\documentclass[draft]{agujournal2019}
\usepackage{url} %this package should fix any errors with URLs in refs.
\usepackage{lineno}
\usepackage[inline]{trackchanges} %for better track changes. finalnew option will compile document with changes incorporated.
\usepackage{soul}
\usepackage{lscape}
\usepackage{amsmath}
\usepackage{amssymb}
\draftfalse

\journalname{AGU Advances}

\begin{document}

%% ------------------------------------------------------------------------ %%
%  Title
%
% (A title should be specific, informative, and brief. Use
% abbreviations only if they are defined in the abstract. Titles that
% start with general keywords then specific terms are optimized in
% searches)
%
%% ------------------------------------------------------------------------ %%

\title{Distant formation and differentiation of outer main belt asteroids and carbonaceous chondrite parent bodies}

%% ------------------------------------------------------------------------ %%
%
%  AUTHORS AND AFFILIATIONS
%
%% ------------------------------------------------------------------------ %%

% Authors are individuals who have significantly contributed to the
% research and preparation of the article. Group authors are allowed, if
% each author in the group is separately identified in an appendix.)

% List authors by first name or initial followed by last name and
% separated by commas. Use \affil{} to number affiliations, and
% \thanks{} for author notes.
% Additional author notes should be indicated with \thanks{} (for
% example, for current addresses).

% Example: \authors{A. B. Author\affil{1}\thanks{Current address, Antartica}, B. C. Author\affil{2,3}, and D. E.
% Author\affil{3,4}\thanks{Also funded by Monsanto.}}

\authors{H. Kurokawa\affil{1}, T. Shibuya\affil{2}, Y. Sekine\affil{1}, B. L. Ehlmann\affil{3,4}, F. Usui\affil{5,6}, S. Kikuchi\affil{2}, and M. Yoda\affil{7,1}}

% \affiliation{1}{First Affiliation}
% \affiliation{2}{Second Affiliation}
% \affiliation{3}{Third Affiliation}
% \affiliation{4}{Fourth Affiliation}

\affiliation{1}{Earth-Life Science Institute, Tokyo Institute of Technology, 2-12-1 Ookayama, Meguro-ku, Tokyo 152-8550, Japan}
\affiliation{2}{Super-cutting-edge Grand and Advanced Research (SUGAR) program, Institute for Extra-cutting-edge Science and Technology Avant-garde Research (X-star), Japan Agency for Marine-Earth Science and Technology (JAMSTEC), 2-15 Natsushima-cho, Yokosuka 237-0061, Japan}
\affiliation{3}{Division of Geological and Planetary Sciences, California Institute of Technology, Pasadena, California 91104, USA}
\affiliation{4}{Jet Propulsion Laboratory, California Institute of Technology, Pasadena, California 91125, USA}
\affiliation{5}{Institute of Space and Astronautical Science, Japan Aerospace Exploration Agency, 3-1-1 Yoshinodai, Chuo-ku, Sagamihara, Kanagawa, Japan}
\affiliation{6}{Center for Planetary Science, Graduate School of Science, Kobe University, 7-1-48 Minatojima-Minamimachi, Chuo-Ku, Kobe, Hyogo 650-0047, Japan}
\affiliation{7}{Department of Earth and Planetary Science, The University of Tokyo, 7-3-1 Hongo, Bunkyo-ku, Tokyo 113-0033, Japan}

%\affiliation{=number=}{=Affiliation Address=}
%(repeat as many times as is necessary)

%% Corresponding Author:
% Corresponding author mailing address and e-mail address:

% (include name and email addresses of the corresponding author.  More
% than one corresponding author is allowed in this LaTeX file and for
% publication; but only one corresponding author is allowed in our
% editorial system.)

% Example: \correspondingauthor{First and Last Name}{email@address.edu}

\correspondingauthor{H. Kurokawa}{hiro.kurokawa@elsi.jp}

%% Keypoints, final entry on title page.

%  List up to three key points (at least one is required)
%  Key Points summarize the main points and conclusions of the article
%  Each must be 100 characters or less with no special characters or punctuation and must be complete sentences

% Example:
% \begin{keypoints}
% \item	List up to three key points (at least one is required)
% \item	Key Points summarize the main points and conclusions of the article
% \item	Each must be 100 characters or less with no special characters or punctuation and must be complete sentences
% \end{keypoints}

% up to three key
% <140 characters for each
\begin{keypoints}
\item Water-rock reactions and their products in asteroids are simulated by hydrological, geochemical, and spectral models.
\item C-type asteroids with ammoniated phyllosilicates possibly formed beyond the NH$_3$ and CO$_2$ snow lines and differentiated.
\item Carbonaceous chondrites can originate from rock-dominated inner cores of the differentiated bodies.
\end{keypoints}

%% ------------------------------------------------------------------------ %%
%
%  ABSTRACT and PLAIN LANGUAGE SUMMARY
%
% A good Abstract will begin with a short description of the problem
% being addressed, briefly describe the new data or analyses, then
% briefly states the main conclusion(s) and how they are supported and
% uncertainties.

% The Plain Language Summary should be written for a broad audience,
% including journalists and the science-interested public, that will not have 
% a background in your field.
%
% A Plain Language Summary is required in GRL, JGR: Planets, JGR: Biogeosciences,
% JGR: Oceans, G-Cubed, Reviews of Geophysics, and JAMES.
% see http://sharingscience.agu.org/creating-plain-language-summary/)
%
%% ------------------------------------------------------------------------ %%

%% \begin{abstract} starts the second page

\begin{abstract} %<250 words
Volatile compositions of asteroids provide information on the Solar System history and the origins of Earth's volatiles. Visible to near-infrared observations at wavelengths of $<2.5\ {\rm \mu m}$ have suggested a genetic link between outer main belt asteroids located at $2.5$--$4\ {\rm au}$ and carbonaceous chondrite meteorites (CCs) that show isotopic similarities to volatile elements on Earth. However, recent longer wavelength data for large outer main belt asteroids show $3.1\ {\rm \mu m}$ absorption features of ammoniated phyllosilicates that are absent in CCs and cannot easily form from materials stable at those present distances. Here, by combining data {collected} by {the} AKARI space telescope and hydrological, geochemical, and spectral models of water-rock reactions, we show that the surface materials of asteroids having $3.1\ {\rm \mu m}$ absorption features and CCs can originate from different regions of a single, water-rock{-}differentiated parent body. Ammoniated phyllosilicates form within the water-rich mantles of the differentiated bodies containing NH$_3$ and CO$_2$ under high water-rock ratios ($>4$) and low temperatures ($<70^\circ$C). CCs can originate from the rock-dominated cores, that are likely to be preferentially sampled as meteorites by disruption and transport processes. Our results suggest that multiple large main belt asteroids formed beyond the NH$_3$ and CO$_2$ snow lines (currently $>10$ au) and could be transported to their current locations. Earth's high hydrogen to carbon ratio may be explained by accretion of these water-rich progenitors. % 226 words
\end{abstract}

\section*{Plain Language Summary} %<200 words
Small bodies record {the} Solar System history and how planets formed. Outer main belt asteroids are thought to be composed of bodies similar to carbonaceous chondrites (CCs) – meteorites which possibly sourced Earth’s highly volatile elements. However, recent spectral observations suggest the presence of ammonia-bearing clays on several large asteroids. Ammonia-bearing clays are absent in {CCs}, and cannot {easily} form from materials stable in the current asteroid belt. To understand the conditions {necessary} to form minerals in those asteroids and CCs, we perform{ed} hydrological and geochemical modeling of water-rock reactions in these bodies. Synthetic spectra {were} computed for the model mineral assemblages and compared with asteroid observations {using} a space telescope. We {found} that surface minerals of outer main belt asteroids, including ammonia-bearing clays, form from starting materials containing NH$_3$ and CO$_2$ under water rich and low temperature conditions, while CC minerals form under water-poor conditions. We propose that multiple large outer main belt asteroids formed at distant orbits and differentiated to form different minerals in water-rich mantles and rock-dominated cores. Asteroid observations are looking at the product of water-rich mantles, {whereas} CCs preferentially sample the rocky cores. This scenario may also explain the elemental composition of Earth’s volatiles. %200 words

%% ------------------------------------------------------------------------ %%
%
%  TEXT
%
%% ------------------------------------------------------------------------ %%

%%% Suggested section heads:
% \section{Introduction}
%
% The main text should start with an introduction. Except for short
% manuscripts (such as comments and replies), the text should be divided
% into sections, each with its own heading.

% Headings should be sentence fragments and do not begin with a
% lowercase letter or number. Examples of good headings are:

% \section{Materials and Methods}
% Here is text on Materials and Methods.
%
% \subsection{A descriptive heading about methods}
% More about Methods.
%
% \section{Data} (Or section title might be a descriptive heading about data)
%
% \section{Results} (Or section title might be a descriptive heading about the
% results)
%
% \section{Conclusions}

%\section{= enter section title =}
%Text here ===>>>

\newpage

\section{Introduction}

Volatile compositions of asteroids record Solar System history and the origins of Earth's volatiles. The link between C-complex {bodies} and carbonaceous chondrite meteorites (CCs) has been suggested for a long time {due to} similarities in their spectral properties including the spectral slope, low albedo, and sharp $3\ {\rm \mu m}$ absorption feature which is characteristic {for} hydrous minerals, centered at $\sim 2.7\ {\rm \mu m}$ \cite{Trigo+2014,Rivkin+2015}. CCs show isotopic similarities {to} highly volatile elements {on} Earth \cite{Dasgupta+Grewal2019}, {indicating} CCs (and thus C-complex asteroids) as the origins of Earth's volatiles. The molybdenum isotope dichotomy between CCs and non-carbonaceous meteorites (NCs) suggests their formation from distinct reservoirs \cite{Kruijer+2017}. Large-scale dynamical evolution has been proposed to have delivered C-type bodies from the outer Solar System, possibly near or beyond the orbits of giant planets \cite<e.g.,>[]{Walsh+2011}. Volatile transport in the Solar System could take place also in the form of icy pebbles \cite{Sato+2016,Nara+2019} and fragments of hydrated bodies \cite{Trigo+2019}.

However, a few large asteroids in the outer main belt possess additional absorption features at $3.1\ {\rm \mu m}$ \cite{King+1992,Rivkin+2006,Milliken+Rivkin2009,Takir+Emery2012}, which are not {observed} in CCs. Several phases have been proposed as the sources of {these} absorption features. The $3.1\ {\rm \mu m}$ {absorption of} 1 Ceres was attributed to ammoniated phyllosilicates or brucite \cite{King+1992,Milliken+Rivkin2009} before the arrival of {the} Dawn spacecraft. In situ observations by Dawn later confirmed that ammoniated phyllosilicates are widespread on {the surface of} Ceres \cite{DeSanctis+2015} (Figure \ref{fig:spectra}b). The $3.1\ {\rm \mu m}$ {absorption} feature {of} 24 Themis was attributed to water ice \cite{Campins2010,Rivkin+Emery2010}. Though several bodies have been classified as Ceres- or Themis-type by using ground-based observations \cite{Takir+Emery2012,Rivkin+2019}, the dominant cause {for} the $3.1\ {\rm \mu m}$ {absorption} features of large outer main belt asteroids has not been fully understood. Moreover, a fundamental question {of} why ammoniated phyllosilicates are present at least in some asteroids and absent in {the} CC collection has not been elucidated. As secondary minerals are the products of water-rock reactions in asteroid precursors (icy planetesimals), when their internal temperatures ($T$) are high enough to melt water ice, the difference in mineral assemblages should reflect that of water-rock reaction conditions. {Water-rock reactions are} also responsible for lithification of bodies initially accreted as porous aggregates by forming secondary aqueous alteration minerals \cite<e.g.,>[]{Rubin+2007}, in concert with compaction due to impacts \cite<e.g.,>[]{Bischoff+2006,Blum+2006,Trigo+2006,Rubin2012,Beitz+2016,Tanbakouei+2020}.

Recently, the AKARI infrared space telescope obtained $2.5$--$5.0\ {\rm \mu m}$ reflectance spectra of large main belt asteroids (mostly $>100$ km), which {were} free from the telluric absorption in prior ground-based data \cite{Usui+2019}. AKARI revealed that $3.1\ {\rm \mu m}$ absorption features are more common than previously thought (Figure \ref{fig:spectra}b and Supplementary Figure 1). Detailed analysis of the $3.1\ {\rm \mu m}$ features {observed} in AKARI spectra to identify the possible phases has not been performed yet.

We aim to understand the source of $3.1\ {\rm \mu m}$ absorption features in outer main belt asteroid spectra and the reason why the material is present only in {the above-mentioned} asteroids and absent in CCs. 
Spectral analysis of AKARI data {is} performed, which leads to a conclusion that $3.1\ {\rm \mu m}$ absorption features are mostly consistent with ammoniated phyllosilicates (Section \ref{sec:observations}). We perform geochemical modeling of water-rock reactions to show ammoniated phyllosilicates and more CC-like mineralogy form under high and low water-rock ratio{s} (W/R, {here defined as an effective mass ratio that tracks the total mass of water to which a unit mass of rock is exposed in the course of aqueous alteration}) conditions, respectively (Section \ref{sec:geochemical}). Spectral modeling of model mineral assemblages is also performed to compare the synthetic spectra to observations (Section \ref{sec:spectra}). We then introduce hydrological modeling of water circulation in asteroids to demonstrate that different W/R {can be} sustained within a single body, namely, in the water-rich mantle and the rock-dominated core of an icy planetesimal (Section \ref{sec:hydrology}). Finally, we discuss the formation and evolution scenario for C-complex asteroids and CC parent bodies and its implications in Section \ref{sec:discussion}, and conclude in Section \ref{sec:conclusions}.

\section{Spectral analysis of asteroid observations {using} AKARI}
\label{sec:observations}

\subsection{AKARI data}
\label{subsec:model_data}

We compared the spectra of C-complex asteroids and our model, especially at $2.7\ {\rm \mu m}$ and $3.1\ {\rm \mu m}$, to constrain conditions of water-rock reactions {for forming} their surface minerals. 
D- and T-type asteroids are also considered because {a} spectral link to ungrouped CC Tagish Lake has been proposed \cite{Hiroi+2001}, though a recent study over a broad spectral range does not support the idea \cite{Vernazza+2013}.
AKARI infrared space telescope obtained the reflectance spectra of 66 large main belt {asteroids}, 25 of which are C-complex and D- and T-type asteroids \cite{Usui+2019}.
We excluded the samples {with unreliable} spectra at any of the absorption bands we are interested {in, following} the criteria {set by} \citeA{Usui+2019} (the uncertainty in the reflectance to be $<10\%$), and used the remaining 19 C-complex, 1 D-type, and 1 T-type asteroids (Supplementary Figure 1). %Correlation plots for their properties are shown in Supplementary Figure 3, which were used to discuss the possible contribution of water ice to $3.1\ {\rm \mu m}$ absorption.
We used Bus-DeMeo taxonomy as summarized by \citeA{Hasegawa+2017}.

When available, the spectra of asteroids {acquired using} AKARI were compared with {the} ground-based telescope (NASA Infrared Telescope Facility, here after IRTF) data \cite{Rivkin+2003,Rivkin+2015ch,Rivkin+2019,Takir+Emery2012}, as well as Dawn data for Ceres \cite{Ciarniello+2017} (Supplementary Figure 2). 
Because of telluric absorption, IRTF data are not available for {the} $2.7\ {\rm \mu m}$ absorption.
Carbonate features at $3.4\ {\rm \mu m}$ are potentially important to constrain the conditions of water-rock reactions. However, we found deviation in this wavelength range between AKARI and IRTF data, possibly due to the contamination of neighboring stars, insufficient background subtraction, or incomplete thermal flux removal \cite{Usui+2019,Rivkin+2019}.
Therefore, spectral comparison to the model results {focused} on $2.7$ and $3.1\ {\rm \mu m}$ absorption features. We discuss implications for future observations to test our modeling with carbonate features (Section \ref{subsec:discussion_future}).

In order to discuss the possible contribution of brucite to $3.1\ {\rm \mu m}$ absorption {while observing} its $2.4\ {\rm \mu m}$ absorption, we combined the reflectance spectra {acquired using} AKARI \cite{Usui+2019} and ground-based observations \cite{Bus+2002,Demeo+2009,Takir+Emery2012} for $>2.5\ {\rm \mu m}$ and $<2.5\ {\rm \mu m}$, respectively. 
Data were compiled by \citeA{Hasegawa+2017}. 

\subsection{Methods}
\label{subsec:model_band}

AKARI, Dawn, and model (see Section \ref{subsec:model_spectra}) spectra at 2.6--3.6 ${\rm \mu m}$ were fitted with 16th-order polynomials, and IRTF spectra at 2.9--3.6 ${\rm \mu m}$ were fitted with 8th-order polynomials to {smoothen} the noise, following \citeA{Rivkin+2019}.
Then, we calculated the absorption band depths, which are defined as {follows} \cite{Usui+2019}:
\begin{equation}
    D = \frac{R_c-R_\lambda}{R_c},
\end{equation}
where $R_c$ is the continuum reflectance {obtained} by {the} interpolation of two local maximums across the absorption feature, and $R_\lambda$ is the reflectance at the peak of absorption.
When multiple absorption bands {overlapped}, we {drew} a common continuum to measure their depths (Supplementary Figure 3).

For error evaluation, we generated 10$^6$ different pseudo-observation spectra for each asteroid spectrum (taken either by AKARI, IRTF, or Dawn) by adding Gaussian noise. Standard deviation of the Gaussian function {was assigned} by observational uncertainties reported in previous studies \cite{Usui+2019,Takir+Emery2012,Rivkin+2003,Rivkin+2015ch,Rivkin+2019}. Spectral fitting and band analysis were performed for each pseudo-observation spectrum. We computed the median and 1-$\sigma$ (more precisely, 68\% confidence interval) values by combining the obtained 10$^6$ absorption-band properties (center, depth, and width). When several IRTF spectra were available for a single asteroid, we computed the median and 1-$\sigma$ values for each of spectrum and then {calculated the average}.\\

\subsection{Results}
\label{subsec:results_AKARI}

We first performed {an} analysis of the $3.1\ {\rm \mu m}$ band shapes in asteroid spectra {acquired using} AKARI \cite{Usui+2019}, as performed to distinguish possible phases on Themis and Ceres in previous studies \cite{Rivkin+2015,DeSanctis+2016}. We found clear detection (1-$\sigma$) for {ten} bodies plus possible detection (band depths are nonzero at their median values but 1-$\sigma$ error bars range to zero) for {three} (Figures \ref{fig:31um} and Supplementary Table 1). Our spectral analysis of band centers {showed} that water ice-coating, whose absorption is centered at $\sim3.08\ {\rm \mu m}$, {was} not responsible for these 3.1 ${\rm \mu m}$ {absorption} features {in most cases} but rather that Ceres-like ammoniated phyllosilcates with band centers $<3.08\ {\rm \mu m}$ \cite{Ehlmann+2018} {were consistent with the observed band centers} (Figures \ref{fig:spectra}b and \ref{fig:31um}). The observed absorption widths {were} also consistent with ammoniated phyllosilicates (0.2--0.3 ${\rm \mu m}$) rather than water ice ($>0.4\ {\rm \mu m}$). Among the 13 bodies, only {one} (24 Themis) {was} more consistent with water ice, and {two others} (52 Europa and 128 Nemesis) {could} be explained by {either} ammoniated phyllosilicates {or} water ice, {or both}.

Furthermore, AKARI spectra {showed} clustering into two {groups} in the $2.7$ and $3.1\ {\rm \mu m}$ absorption depth plot (Figure \ref{fig:features}). Bodies with deep ($>20$\%) $2.7\ {\rm \mu m}$  absorption {did} not show $3.1\ {\rm \mu m}$ absorption, but those with shallow ($<20$\%) absorption {did}. This suggests that the abundances of two phases responsible for {$2.7\ {\rm \mu m}$ and $3.1\ {\rm \mu m}$} features are controlled by a single parameter, otherwise we {would} not see the correlation. As we show in Section \ref{subsec:results_geochemical}, serpentine and ammoniated saponite are the candidates, as their abundances are controlled by a single parameter: {the reaction} W/R. We further show that our geochemical and spectral models can reproduce the trend seen in observations (Section \ref{subsec:results_spectra}).

We note that \citeA{Rivkin+2019} also analyzed $3.1\ {\rm \mu m}$ band centers in asteroid spectra with ground-based observation data. Their conclusions {are consistent with ours in that $3.1\ {\rm \mu m}$ absorption features are more common on large outer main belt asteroids than previously thought}, but they {did} not find {a} hiatus in the distribution to distinguish Ceres-like (ammoniated phyllosilicates) and Themis-like (water ice) objects. We utilized space telescope data which are free from {telluric} absorption around $3\ {\rm \mu m}$, analyzed the band widths as well as the center positions for a larger number of asteroid samples, and concluded that the majority of them {were} Ceres-like.

Because AKARI spectra contain wavy patterns possibly due to the contamination of neighboring stars or insufficient background subtraction \cite{Usui+2019}, we also utilized independent spectra {acquired using} IRTF \cite<e.g.,>[]{Takir+Emery2012,Rivkin+2019} and Dawn data for Ceres \cite{Ciarniello+2017}. Comparison of $3.1\ {\rm \mu m}$ absorption {showed} quantitative consistency (Supplementary Figure 4 and Supplementary Table 1). Most importantly, {two clusters} were found in both AKARI and IRTF spectra (Supplementary Figure 5 and Supplementary Table 1). IRTF spectra {showed shallower} absorption than AKARI spectra, which might be caused by wavy patterns in AKARI spectra \cite{Usui+2019} or {the} artificial cutoff at $2.9\ {\rm \mu m}$ in IRTF spectra.
Ceres spectra {acquired by} Dawn {showed} somewhat deeper absorption than AKARI and IRTF.
We do not discuss the cause of quantitative difference further, but the origin of the the qualitative trend {observed} in both AKARI and IRTF data is discussed (Section \ref{subsec:results_spectra}).

Additionally, the lack of correlation between $3.1\ {\rm \mu m}$ absorption depth and semi-major axis in {the} AKARI data (Supplementary Figure 6) as well as {the} short lifetime of water ice on asteroid surfaces in the main belt \cite{Schorghofer+2008,Beck+2011,Brown2016} does not support water ice as the dominant source of these common features.
We note that \citeA{Takir+Emery2012} and \citeA{Rivkin+2019} reported {the} correlation {between the heliocentric distance and $3\ {\rm \mu m}$ absorption} using a smaller number of samples and data set {acquired using} IRTF.

We also ruled out brucite as a major cause of the $3.1\ {\rm \mu m}$ absorptions on the main belt asteroids observed {using} AKARI. AKARI spectra combined {with} ground-based observation data \cite{Hasegawa+2017} did not {show} the $2.45\ {\rm \mu m}$ absorption (Supplementary Figure 7), which {suggested the absence of brucite}. The AKARI data thus indicate more Ceres-like, ammonium phyllosilicate-bearing asteroids than previously identified \cite{Takir+Emery2012,Rivkin+2019}, {increasing the need} to explain the origin of ammonium, as NH$_3$ ice is unstable at those distances in the present Solar System.
%our chemical model does not predict substantial brucite formation (Supplementary Figure 2) 

Paradoxically, {despite} the fact that Ceres is the largest asteroid and several other large asteroids exhibit similar spectral properties, ammoniated phyllosilicates have not been found in CCs, such that the $3.1\ {\rm \mu m}$ absorption is not seen (Figures \ref{fig:spectra}c and \ref{fig:features}). 
Though long-term collisional evolution likely affected their mineralogy through gardening and mixing \cite{Bischoff+2006,Trigo2015,Beitz+2016,Nittler+2019,Tanbakouei+2020} and, consequently, infrared reflectance spectra, the common ammoniated features on many large asteroids, that dominate the mass of the asteroid belt and are thought to be primitive remnants of planetesimal formation \cite{Bottke+2005,Bottke+2015}, require an explanation for {their differences with} CCs from their original water-rock reaction conditions, such as starting material {composition}, W/R, and $T$ of alteration.

\section{Geochemical modeling}
\label{sec:geochemical}

\subsection{Methods}

Water-rock reactions in the molten stages of icy planetesimals were modeled. 
In the thermodynamic calculations, a mean composition of CV chondrites was assumed for the initial bulk rock \cite{Pearson+2006,Henderson+2009,Clay+2017} (Supplementary Table 2). Minor {amounts} of carbon, nitrogen, and chlorine {were} also included.
In contrast to previous studies \cite{Zolensky+1989,Rosenberg+2001,Schulte+Shock2004,McAlister+Kettler2008,Zolotov2012,Castillo+2018}, our model explicitly {computed} the abundance of NH$_4$-bearing phyllosilicates and solid solution, the former of which is the key mineral responsible for $3.1\ {\rm \mu m}$ absorption \cite<e.g., Mg-saponite,>[]{Ehlmann+2018}, and we analyzed its stability for a wide range of {temperatures} and W/R. 

{W/R, a driving parameter in our simulation was explored for} the range of 0.2$<$W/R$<$10 (the mass ratio). For the initial fluid, three cases were assumed (Table \ref{tab:model}). 
Whereas Case 1 {simulated} reactions of pure H$_2$O and the rock component as typically assumed for CC's parent bodies, Case 3 {contained} NH$_3$ and CO$_2$ in addition to H$_2$O, assuming accretion beyond {the} NH$_3$ and CO$_2$ snow lines. 
CO$_2$ {concentrations} {were} 0, 1 and 10 mol\% (Cases 1–3, {respectively}) relative to H$_2$O while the latter two cases also {included} NH$_3$ (0.5\%) and H$_2$S (0.5\%). NH$_3$ and CO$_2$ were included to form ammoniated phyllosilicates and carbonates. H$_2$S was also considered because its snow line is close to those of NH$_3$ and CO$_2$ \cite{Okuzumi+2016} and it sources sulfur to form sulfur-bearing minerals.
The model composition of Case 3 was {established} by considering comets as a reference. There is no significant bulk compositional variation between comets from the Kuiper belt or from the Oort cloud, and the abundances of CO$_2$, NH$_3$, {and} H$_2$S relative to H$_2$O range 2--30\%, 0.3--1\%, and 0.3--1\%, respectively \cite<the number of samples is about 10,>[]{Mumma+2011,bockelee+Biver2017,Rubin+2020}. We assumed roughly averaged values in Case 3. Case 2 {was the} intermediate between the two with the reduced {CO$_2$} abundances of at $1\%$ relative to H$_2$O.

The equilibrium temperature and pressure were assumed to be 0, 100, 200, 300 and 350$^\circ$C, and saturated vapor pressure of water. 
The temperatures used by our chemical equilibrium model correspond to the quenching {temperature}, i.e., the temperature at which water is no longer available for the reaction. 
After that time, the reactions {are halted} or slowed enough to preserve the minerals.
Here we do not specify the mechanisms of quenching, which include freezing due to secular cooling (quenching at the freezing point), freezing due to abrupt temperature decease, and abrupt water loss (quenching at the temperature). The latter two {might} be caused by events such as water exhalation \cite{Young+2001,Young+2003} and catastrophic disruption \cite<e.g.,>[]{Jutzi+2010}. We note that complete consumption of water by reactions does not occur {at} W/R $>$ 0.2. Considering serpentinization, the dominant water-rock reaction for low W/R, the critical value {below which water consumption is complete} is W/R = 0.17.

In the calculations, pyrene was considered as a representative of polycyclic aromatic hydrocarbon typically found in insoluble organic matter (IOM) while C1 compounds, except {for} CH$_4$, were included as soluble species \cite{Zolotov2012}. 
Because formation of some minerals (e.g., chlorite, olivine, clinopyroxene, garnet, mica, tephroite) is kinetically very slow at low temperatures, the precipitation of these minerals was suppressed in the thermodynamic calculations at temperatures lower than 100$^\circ$C. 

In the water-chondrite reactions, molecular hydrogen is generated due to reduction of water by metal iron and FeO in chondrites, which potentially elevates $f_{\rm H_2}$ of {the} fluid above H$_2$ saturation level in some cases. 
Therefore, it was assumed that $f_{\rm H_2}$ of fluid {did} not exceed water pressure ($P_{\rm H_2O}$). When $f_{\rm H_2}>f_{\rm H_2O}$, the {H$_2$} gas phase {formed} and {was} assumed to escape from the system.

The thermodynamic calculations of water-chondrite reactions were conducted {using the} EQ3/6 computer code \cite{Wolery+Jarek2003}. 
The thermodynamic database required for the calculations was generated {using} SUPCRT92 \cite{Johnson+1992} with thermodynamic data for {minerals}, aqueous species and complexes \cite{Helgeson+1978,Shock+Helgeson1988,Shock+Koretsky1995,Shock+1989,Shock+1997,Sverjensky+1997,Mccollom+2009}. 
Thermodynamic parameters for a series of smectites were estimated using the procedure {published by} \citeA{Wilson+2006}.

\subsection{Results} 
\label{subsec:results_geochemical}

%We next constrain conditions of water-rock reactions to form ammoniated phyllosilicates seen in spectra of large outer main belt asteroids. 
We {quantified} the mineral assemblages {at} chemical equilibrium for various W/R ($0.2$--$10$), $T$ ($0$--$350\ {\rm ^\circ C}$), and NH$_3$ and CO$_2$ abundances, in which NH$_4$-bearing phyllosilicates {were} explicitly modeled. The key results are summarized in Figure \ref{fig:mineral_abundance} and all results are shown in Supplementary Figure 8.

Hydrous minerals {represent} the dominant phase {at} $T<300^\circ$C, whereas anhydrous minerals become dominant {at} $T$ = 350$^\circ$C (Figure \ref{fig:mineral_abundance}). The dominant hydrous phase {was} serpentine at lower W/R (e.g., W/R $<4$ for Case 2), saponite at higher W/R (e.g., W/R $>4$ for Case 2), and chlorite at higher $T$ (e.g., $T$ =  350$^\circ$C). 

Ammoniated saponite {appeared} only when NH$_3$-ice {was present} in the starting material (Cases 2 and 3), at high W/R ($>4$ and $>0.4$ for Cases 2 and 3, respectively), and {at} $T\sim0\ {\rm ^\circ}$C (Figure \ref{fig:mineral_abundance}).
Lower W/R {led} to higher pH of rock-buffered fluids where the dominant hydrous phase {was} serpentine rather than saponite. 
Furthermore, at pH $>12$, NH$_4$-bearing saponite {could not form} {due to the} conversion of NH$_4^+$ to NH$_3$ {and} H$^+$ in the porewater.
With {the} increasing temperature, NH$_4$-saponite {was} replaced with Fe-saponite at around 70$^\circ$C (Figure \ref{fig:Tdependence}). This {happened} because the low pH at high $T$ {led} to high concentrations of divalent ions, including ferrous ion, in fluids, which, in turn, stabilizes Fe(II) in the interlayer rather than NH$_4$-bearing saponite, consistent with a previous study \cite{Neveu+2017}.
The relative abundance of saponite and serpentine as a function of W/R {was} consistent with previous studies \cite{Neveu+2017,Castillo+2018}.

Carbonates {appeared} at high W/R and {became} more abundant as the CO$_2$ abundance increases from Cases 1 to 3. 
As W/R {got} lower, organics {became} stable instead of carbonates due to reduced conditions, caused by H$_2$ production and its high concentration.

\section{Spectral modeling}
\label{sec:spectra}

\subsection{Methods}
\label{subsec:model_spectra}

The synthetic infrared reflectance spectra of model mineral assemblages remaining after the water-rock reactions were calculated by adopting the radiative transfer model for granular surfaces \cite{Hapke2012} and compared with the AKARI data to constrain the conditions of water-rock reactions which formed minerals on large outer main belt asteroids and in CCs. Because the Hapke model does not fully reproduce the reflectance of dark mineral assemblages from mixing of endmembers \cite<see,>[and references therein]{Kurokawa+2020}, our purpose to use the spectral model {was} limited {in demonstrating} that our model {could} reproduce the observed trend in absorption depths, and we {did} not attempt to fully reproduce asteroid spectra (albedo, spectral slope, etc.) and constrain grain sizes. Because the reflectance spectra obtained {using} AKARI {were} given in the form of geometric albedo (also called physical albedo, which is the ratio of its actual brightness as {observed} from the light source to that of an Lambertian disk with the same cross-section), our model computed the geometric albedo $A_p$ from the weighted average of the single scattering albedos (SSAs) of endmembers, $w(\lambda) = \sum f_i w_i(\lambda)$, where $\lambda$ is the wavelength and $f_i$ is the fractional relative cross section of component $i$. 
Assuming isotropic scatters, the geometric albedo $A_p(\lambda)$ is approximated by \cite{Hapke2012},
\begin{equation}
    A_p(\lambda) = 0.49 r_0(\lambda) + 0.196 r_0(\lambda)^2,
\end{equation}
with {an} accuracy better than 0.4\%, where $r_0(\lambda) = (1-\gamma(\lambda))/(1+\gamma(\lambda))$ is the diffusive (bihemispherical) reflectance, and $\gamma(\lambda) = \sqrt{1-w(\lambda)}$.
The endmember SSAs $w_i$ were calculated through the refractive indices $n$ and $k(\lambda)$ from the endmember reflectance spectra  \cite<more precisely, the reflectance factor,>[]{Hapke2012} $r_i(\lambda)$ taken from RELAB and the literature (Supplementary Table 3). 
The real refractive index $n$ was assumed to be independent from the wavelength for the range considered here and was taken from the literature.
The imaginary refractive index $k(\lambda)$ was calculated from the reflectance data and the given $n$ value {via the} inversion of {the} Hapke model, as summarized by \citeA{Lapotre+2017}.
We assumed that grains {were} spherical, {and the grain diameter for} isotropic scatters {was} $d$. 
This modeling approach has been used and validated with natural samples and lab mixtures {in} previous studies \cite{Lucey1998,Lawrence+Lucey2007,Lapotre+2017}.

Reflectance data for the wavelength range 2.5--4.0 ${\rm \mu m}$ {were} sometimes not found for minor minerals (NH$_4$-beidellite, tremolite, tephroite, sylvite, and Mn(OH)$_2$). 
We assumed the same refractive constants with NH$_4$-saponite for NH$_4$-beidellite, and neglected the optical effects of the others.
Because the abundances of those minerals in our model are low, the assumption would not influence the calculated spectra.

We assumed that the grain diameter $d$ {was} the same for all minerals except for IOM for simplification. 
The nominal size {was} $d = 100\ {\rm \mu m}$ for minerals and $d = 0.5\ {\rm \mu m}$ for IOM unless otherwise stated.
The small grain size of IOM was assumed to increase its total cross section, and thus, to make the modeled albedo as low as those {for the} observed asteroids.
It is a well known problem that radiative-transfer models underestimate the influence of dark phases \cite<see,>[and references therein]{Kurokawa+2020}. For this reason previous studies \cite{Rivkin+Emery2010,DeSanctis+2015} assumed a very high abundance of amorphous carbon or magnetite. Our approach is in line with those studies. 
IOM particle sizes in meteorites are as small as assumed here \cite<typically $<1\ {\rm \mu m}$, e.g.,>[]{Alexander+2017} as well as other dark phases in {the} matrix \cite<e.g.,>[]{Buseck+Hua1993} which cause the low reflectance of CCs \cite{Garenne+2016,Kurokawa+2020}.
Because Hapke's SSA model cannot be applied to grain sizes smaller than the considered wavelengths, we {assumed} the same SSA (calculated from {the} measured reflectance) for IOM regardless of the grain size in order to avoid any unphysical behavior.

Thin coating of minerals by water ice has been proposed for the origin of $3.1\ {\rm \mu m}$ absorption on 24 Themis \cite{Rivkin+Emery2010}. To discuss the presence of water ice on asteroids' surfaces to cause $3.1\ {\rm \mu m}$ absorption, we computed linear mixing of the water-ice coating model \cite<>[Supplementary Figure {9}]{Rivkin+Emery2010} and {the} AKARI spectrum of an asteroid (2 Pallas) without $3.1\ {\rm \mu m}$ absorption. The band center and width {were} computed with the band detection method described in Section \ref{subsec:model_band}.

\subsection{Results}
\label{subsec:results_spectra}

Typical model spectra are shown in Figure \ref{fig:spectra}a and all results are shown in Supplementary Figure 10. As expected, the diagnostic secondary minerals {were} seen in modeled spectra: hydrous minerals at $2.7\ {\rm \mu m}$, ammoniated phases at $3.1\ {\rm \mu m}$, and carbonates at $3.4\ {\rm \mu m}$ and $4.0\ {\rm \mu m}$.
All modeled reflectance spectra {showed} a $2.7\ {\rm \mu m}$ OH-absorption. The diagnostic $3.1\ {\rm \mu m}$ absorption feature of ammoniated saponite {appeared} only when NH$_3$-ice {was present} in the starting material (Cases 2 and 3),  at high W/R ($>4$ and $>0.4$ for Cases 2 and 3, respectively), and $T\sim0\ {\rm ^\circ}$C. Carbonate absorptions at $3.4\ {\rm \mu m}$ and $4.0\ {\rm \mu m}$ {were} seen for all modeled spectra, and {got} deeper when starting materials contain more carbon as CO$_2$ ice (Cases 1--3).

Comparison of the absorption depths between {the} model and AKARI observations indicates that high W/R ($>4$ in Case 2), low $T$ ($<70^\circ$C), and fluids containing NH$_3$ and moderate CO$_2$ are required to form the surface mineralogy of several large main belt asteroids showing $3.1\ {\rm \mu m}$ features (Figure \ref{fig:spectra}b). Observed spectra of both AKARI and IRTF samples {showed} clustering of two groups in the $2.7$ and $3.1\ {\rm \mu m}$ absorption depth plot (Figure \ref{fig:features}, Section \ref{subsec:results_AKARI}). The clustering can be explained {via} the change {in} the dominant hydrous phase from serpentine to ammonia-bearing saponite with increasing W/R at $T =  {\rm 0^\circ C}$ with a single bulk volatile composition (either Case 2 or 3). The observed but relatively shallow $3.4\ {\rm \mu m}$ absorption {observed} in Ceres global average spectrum {acquired using} Dawn {was} consistent with an initial bulk composition depleted in CO$_2$, compared to comets (Case 2 vs. Case 3, Figures \ref{fig:spectra}a and \ref{fig:spectra}b). 
In contrast, the bodies not showing the $3.1\ {\rm \mu m}$ absorption and CCs (Figure \ref{fig:spectra}c) {were} reproduced either {at} lower W/R or {at} higher $T$ ($>70^\circ$C, Section \ref{subsec:results_geochemical}), which {led} to the absence of NH$_4$-saponite even though the starting material contained NH$_3$ ice. 

Supplementary Figure 11 shows the dependence of absorption band depths on the grain sizes.
Decreasing grain size $d$ {led} to brighter overall reflectance and shallower absorption depths. 
Varying grain size covers a part of variation in observed absorption depths.
Varying grain size from 200 $\mu$m down to 50 $\mu$m still {showed} qualitative {similarity} to {the} observations. This is important to note, as we do not know {the} actual grain sizes of asteroid surfaces to which we compared our model spectra. Thanks to the Dawn mission, Ceres is the best investigated body among them. However, estimates for its surface grain size varies between studies from a few $\mu$m to $\sim$100 $\mu$m \cite{Prettyman+2017,Li+2019,Raponi+2019,Kurokawa+2020}. Particles suspended in the water-rich mantle would be $\mu$m size initially \cite{Neveu+2015,Bland+Travis2017,Travis+2018}, but extensive aqueous alteration may cause grain growth {\cite<e.g.,>[]{Jones+Brearley2006}}.

These results suggest that secondary minerals of asteroids showing $3.1\ {\rm \mu m}$ absorption and of CCs formed under different W/R conditions (W/R $>4$ and $<4$, respectively, in Case 2). Below, we hypothesize that such distinct conditions might be present in the water-rich mantle and porous rocky core of a single, water-rock differentiated body, and constrain the conditions to maintain the difference in W/R by simulating water flow in an icy planetesimal. We note that our term \textquotedblleft differentiation\textquotedblright \ should not be confused with rock-metal differentiation. We discuss water-rock differentiation owing to segregation of rock grains to form a rock-dominated core. 

\section{Hydrological modeling}
\label{sec:hydrology}

\subsection{Methods}

In order to investigate whether two different W/R {values} (higher and lower than 4 in the water-rich mantle and the rock-dominated core in Case 2) {are} possible within {the} same body, we performed hydrological simulations for the interiors of icy planetesimals. In contrast to previous studies \cite{Travis+Schubert2005,Palguta+2010,Bland+Travis2017,Travis+2018}, which modeled the entire body to demonstrate that compositional variation within a single planetesimal is possible, we {focused} on the velocity of fluid flow in the outer part of the porous rocky core. In this part, the temperature decreases toward the surface, which in turn could have potentially caused fluid convection in the core \cite<e.g.,>[]{Young+2003}. 
We employed a fluid-flow simulator {called} GEneral purpose Terrestrial fluid-FLOW Simulator \cite<GETFLOWS,>[]{Tosaka+2000}. 
GETFLOWS is a finite difference fluid-flow simulator that can solve movement of both surface water and subsurface fluid flows simultaneously. 
GETFLOWS can treat multi-phase components (i.e., water, gas, and solidified rock). 
Subsurface fluid flow {was} calculated based on Darcy's equation and mass, momentum, and energy conservations. 
Given the low gravity of planetesimals, we employed vertically-variable gravitational acceleration, $g(r)$, based on the following equation: $g(r) = g_0\cdot r/R$, where $g_0$ is the gravity at the surface, $r$ is the radial distance from the center, and $R$ is the radius of the planetesimal. 

For simplicity, the computational area was discretized in a two-dimensional, Cartesian grid-block system (total length of 240 km with 120 grids and total depth of $\sim$50 km with 21 grids). 
The use of Cartesian grid and ignoring spherical geometry is justified because we focus on a small domain in a planetesimal, rather than simulating the entire body.
The uppermost grids (height of 10 km) of the grid-block system {were} defined as a water-rich layer without solidified rocks. 
The underlying 20 grids (height of 2 km for one grid) {were} composed of a mixture of fluid and solidified rocks. 
No heat production was assumed throughout the simulations. 
No exchange of fluids and heat with the outside of the grid-block system was also assumed. 

The surface topography {was} based on a digital terrain model for an equatorial region of Ceres \cite<longitude 160--190$^\circ$E,>[]{Roatsch+2016}. The results {were} not sensitive to the surface topography, and we adopted {Ceres topography} as an example, as Ceres is the best investigated body among large C-complex asteroids.

The major parameters of our simulations {were} initial core temperature and gravity (i.e., the size of the planetesimal). 
We set the initial temperature of the bottom grids, $T_{\rm rock}$, as 100$^\circ$C or 300$^\circ$C. 
The temperature of the overlying water-rich layer was kept {at} 0$^\circ$C throughout the simulations. 
A linear thermal gradient was assumed for the initial temperature profile in the subsurface. 
In our simulations, the rocky core was cooled due to interactions (conduction or convection) with the low-temperature water-rich layer throughout the simulations. 
In reality, temperatures in planetesimals are dependent on the timing of accretion (the amount of radioactive elements such as $^{26}$Al), heat transport, and of course the time since the accretion, and thus variable temperatures between different bodies are possible. The above mentioned temperatures in our hydrological model {were} chosen to cover the temperature range where hydrous minerals are found to be stable in our geochemical modeling (Section \ref{subsec:results_geochemical}) and the range {in} which various classes of CCs have experienced \cite{krot1998a,Krot+1998b,Krot+2006,Brearley2006}.

As for gravity, we considered planetesimals with radius $R$ of 200 or 500 km with the bulk density of {2000} kg/m$^3$. 
In our spectral samples, only Ceres size is comparable to the latter, and the others ($R<\sim 250$ km) can be approximated by the former.

We used the hydrological properties (viscosity, density, thermal expansion rate, and capillary pressure) of pure water \cite{Tosaka+2000} for fluids both in the porous core and in the water-rich mantle (the uppermost grids). In reality, the fluid contains fine grains and solutes \cite<e.g.,>[]{Travis+2018}, but their effects are expected to be minor. Assuming W/R $>4$ (the mass ratio) as suggested from our geochemical model to form ammoniated phyllosilicates for the fluid, the increase in viscosity due to entrained fine particles are less than a factor of $\sim$2 \cite<particle volume concentration $<$0.1 in Figure 3 of>[]{Travis+2018}. The difference in viscosity between pure water and brine is less than 10\% {for the low solute concentrations considered in our geochemical model \cite<e.g.,>[]{Kargel+1991,Ozdemir+2007}}. The freezing-point depression due to the solutes may affect the flow near the freezing point \cite{Travis+2018}, but we {focused} on high-temperature phases where convection {was} intense and thus its influence is not important for our purpose.

The permeability $k_p$ was randomly varied in the subsurface so that the distribution of permeability follows {a} Gaussian distribution with the mean value and standard deviation of 10$^{-12\pm 1}$ m$^2$. 
The mean permeability {was} based on the data for fine sands and those proposed for icy planetesimals \cite{Grimm+McSween1989,Cohen+Coker2000,Coker+Cohen2001,Young+1999,Young+2001,Young+2003,McSween+2002,Travis+Schubert2005}. The permeability depends on the particle size and porosity \cite{Travis+2018} and is lower for CCs \cite{Bland+2009}. The adapted value is likely {an} upper limit, and we expect that the porosity would decrease with time as secondary minerals fills in the pore \cite<as we observe in CCs,>[]{Rubin+2007}. If we assume lower values, flow becomes more sluggish and thus the difference in W/R would be sustained for even higher core temperatures (Section \ref{subsec:results_hydrology}). Thus, our assumption on the permeability is a conservative approach.

The heat capacity, thermal conductivity, and porosity were assumed to be uniform in the subsurface as 1000 J/kg/K, 3.0 W/m/K, and 0.35, respectively. 
We calculated time evolution of fluid temperature and pressure and their movements upon cooling of the porous rocky core. 
The results were analyzed using a data visualization software of ParaView 5.6.1. 

\subsection{Results}
\label{subsec:results_hydrology}

Figure \ref{fig:water_circulation} demonstrates snapshots of vector, velocity, and temperature of fluids in the subsurface {at} different initial core temperatures ($T_{\rm rock}$ = 100 and 300$^\circ$C) and different sizes of planetesimals ($R$ = 200 and 500 km in radius).
Our results suggest that, at the high core temperatures and high gravity, various W/R and temperatures can be achieved within a single icy planetesimal.
Convection of fluids {occurred when} $T_{\rm rock}$ = 300$^\circ$C and $R$ = 500 km (Figure \ref{fig:water_circulation}a). 
Figure \ref{fig:water_circulation}a shows that upwelling of hydrothermal fluids ($\sim$200--250$^\circ$C) {occurred} at {a} high velocity of 0.2--0.4 mm/day (70--150 km/Myr). 
Downwelling of cold water from the water-rich layer also {happened} at velocity of 0.05--0.1 mm/day (20--40 km/Myr). It would take the order of $\sim$1 Myrs for fluids to circulate within a $\sim$100-km-sized planetesimal given the fluid velocity of 0.1 mm/day ($\sim$40 km/Myr). 
Such a fluid circulation between the rocky core and water-rich layer {continued} for a few Myrs upon cooling. 
Due to the fluid circulation, the W/R {was the} highest along the fluid flow paths of the circulation cells.
For instance, effective W/R in the fluid flow path {increased} (e.g., W/R $\sim$ 1) a few times {the local mass ratios} that determined by porosity. 
In contrast, W/R {remained} low (e.g., W/R $\sim$ 0.3--0.5) near the center of the circulation cells because of {the} lack of fluid input from the water-rich layer. 

{In contrast}, at the low core temperatures and/or low gravity, the W/R in the rocky core {remained} low ({the value close to the local mass ratio,} e.g., W/R $\sim$ 0.3--0.5) and distinct from that of the overlying water-rich mantle (e.g., W/R $>4$ to form ammoniated phyllosilicates in Case 2, Section \ref{subsec:results_geochemical}).
No convection {occurred} for $T_{\rm rock}$ = 100$^\circ$C and/or $R$ = 200 km (Figure \ref{fig:water_circulation}b--d). 
Under these initial conditions, fluids in the subsurface {flew only} through limited regions where permeability {was} locally {high}, forming veins in the rocky core. We note that the locations of upwelling and downwelling {varied} as we {changed} the random distribution of the permeability, but the general trend still {held}. Upon cooling and volume reduction of water, inhalations (single pass flow) of fluids {occurred} along with the veins. 
However, typical velocity of fluids at veins {was} 0.01--0.03 mm/day (3--10 km/Myr); accordingly, the W/R at veins would remain low even if inhalation {occurred}. We note that the fluid velocity obtained in our simulations {was} consistent with previous studies \cite<e.g., $\sim$ 10 km/Myr as shown in Figure 2A of>[]{Bland+Travis2017}.

The difference in the modes of fluid flow dynamics (i.e., convection vs. inhalation) can be explained based on the Rayleigh-Darcy number, $Ra$ \cite<e.g.,>[]{Young+2003}. 
The Rayleigh-Darcy number is expressed as follows based on the definition of the Rayleigh number and the Darcy’s law; 
\begin{equation}
    Ra = \frac{\rho \beta \Delta T k_p L g}{\alpha \eta},
\end{equation}
where $\rho$ is the fluid density, $\beta$ is the thermal expansion coefficient of fluid, $\Delta T$ is the temperature, $k_p$ is the permeability, $L$ is the length, $g$ is the gravity, $\alpha$ is the thermal diffusivity, and $\eta$ is the dynamic viscosity of fluid. When $Ra$ is $<\sim$10$^3$, no convection occurs in the system. 
Using the physical properties of pure water and our calculation conditions, $Ra$ is less than $\sim10^3$ for the 200 km planetesimal and/or initial core temperature of 100$^\circ$C. {If a planetesimal is 200 km or smaller, and/or if its core temperature is 100$^\circ$C or less, no convection of pore water occurs.}
{In contrast}, $Ra$ becomes $>10^3$ for the 500-km planetesimal and initial core temperature of 300$^\circ$C. 

To summarize, the difference in W/R between the water-rich mantle and the porous rocky core {was} best maintained when $T_{\rm rock}$ = 100$^\circ$C and/or $R$ = 200 km (typical for our spectral samples). W/R in the core {varied} when $T_{\rm rock}$ = 300$^\circ$C and $R$ = 500 km (Ceres' size). Though we do not fully rule out the latter case, high W/R to form ammoniated phyllosilicates might be achieved even in the core in this case.

\section{Discussion}
\label{sec:discussion}

\subsection{Formation and evolution scenario for outer main-belt asteroids}

Based on the results presented in Sections \ref{sec:observations}--\ref{sec:hydrology}, we propose a new idea that both Ceres-type materials containing ammoniated phyllosilicates and CCs-type materials formed in water-rock differentiated bodies which formed as icy planetesimals beyond the NH$_3$ and CO$_2$ snow lines and thus accreted NH$_3$ and CO$_2$ ice in addition to water ice (Figure \ref{fig:scenario}, Stage 1). Their different mineral assemblages result from water-rock differentiation and consequent vertical gradients of W/R, temperatures, and pH within the single body. Assuming NH$_3$ and CO$_2$ sublimation temperatures {of} 75 and 85 K \cite{Okuzumi+2016}, respectively, and the equilibrium temperature profile in the current Solar System $T_{\rm eq} = 280\ (r/{\rm au})^{-1/2}\ {\rm K}$ (where $r$ is the orbital radius), {the} NH$_3$ and CO$_2$ snow lines are located at 11 and 14 au, respectively.

Parent bodies depleted in CO$_2$ compared to comets can best explain the amount of carbonates {inferred} from $3.4\ {\rm \mu m}$ and $4.0\ {\rm \mu m}$ absorption features {of Ceres} (Section \ref{subsec:results_spectra}). CO$_2$ depletion is expected as a consequence of degassing to space upon heating and melting of icy planetesimals (Figure \ref{fig:scenario}, Stage 2).
In order to quantify CO$_2$ degassing, we performed additional, low-temperature thermodynamic calculations of the H$_2$O-NH$_3$-CO$_2$ mixture with the cometary composition. We found that $90\ {\rm wt.\%}$ of CO$_2$ {was} released from the mixture as gas at 0$\ ^\circ$C, whereas NH$_3$ {was} retained as NH$_4$-hydrate fluid (Supplementary Text S1 and Supplementary Figure 12). 
This preferential loss of CO$_2$ to space {led} to the CO$_2${-}depleted bulk composition relative to cometary ice (Case 2 composition, Section \ref{subsec:results_geochemical}). 

Further heating {induced} water-rock differentiation within bodies (Figure \ref{fig:scenario}, Stage 2). 
While large, millimeter-sized grains (chondrules) {settled} down to form {a} low-W/R interior (porous rocky core), smaller grains (matrix) {remained} suspended by convection in water, forming a high-W/R upper-layer (ice-covered, muddy subsurface ocean) \cite{Neveu+2015,Bland+Travis2017,Travis+2018}.
The upper layer {water} would be eventually lost to space via sublimation, but the observed $3.1\ {\rm \mu m}$ absorption feature of ammoniated phyllosilicates (Section \ref{subsec:results_AKARI}) and high W/R ($>4$) and low temperature ($<70^\circ$C) conditions required to form these ammoniated phases (Section \ref{subsec:results_geochemical}) indicate sufficient retention of subsurface water {during} heating so that these volatiles {participated} in water-rock reactions. 
The estimated lifetime of {water} ice on small bodies at the orbit of the main asteroid belt \cite<$\sim 10^3$--$10^4$ years,>[]{Machida+Abe2010} is much shorter than the period of heating by radioactive decay \cite<$\sim10^6$ years,>[]{Castillo2010}. 
However, the spectral detection of water ice on 24 Themis \cite{Campins2010,Rivkin+Emery2010} and the low bulk density of Ceres suggest that lifetime can be much longer, due to an insulating regolith layer \cite{Schorghofer+2008}.
Freezing prior to significant water loss {led} to the occurrence of the particular minerals formed at $T<70^\circ$C, (namely, NH$_4$-saponite) and at W/R $>4$ (carbonates) in the icy mantle (Figure \ref{fig:scenario}, Stage 3). Alternatively, the prolonged lifetime of an H$_2$O-ice mantle can be naturally explained if the asteroids are delivered to the current location well after the freezing.
After freezing, sublimation and loss of ice to space leave these secondary minerals on asteroid surfaces as lag deposits, causing their spectra around $3.1\ {\rm \mu m}$ being chiefly dominated by ammoniated phyllosilicates (not by water ice, Section \ref{subsec:results_AKARI}).

Some of the observed asteroids {did} not show $3.1\ {\rm \mu m}$ absorption and {resembled} CM/CI chondrites (Section \ref{subsec:results_spectra}). These {objects} might {have experienced} insufficient water-rock differentiation, which {led} to lower W/R even in their upper layers, {inhibiting} ammoniated phyllosilicate formation even though the starting material contained NH$_3$ (e.g., Case 2 with low W/R, Section \ref{subsec:results_geochemical}).
Additionally, water loss prior to freezing due to exhalation \cite{Young+2001,Young+2003} or catastrophic disruption \cite<e.g.,>[]{Jutzi+2010} would leave only secondary minerals that {were} stable at higher $T$ and lower W/R. Therefore, it is possible that these bodies accreted NH$_3$ and CO$_2$ ice but experienced a different evolution pathway. 

Moreover, non-disruptive impacts through the Solar System history influenced asteroid surface compositions by removing and excavating endogenous materials and replenishing impactor materials. Decomposition of hydrous minerals due to impacts is expected to be limited \cite{Wakita+Genda2019}. Cumulative impacts have affected Ceres surface composition, though endogenous NH$_4$-bearing phyllosilicates still remain \cite{Stein+2019,Marchi+2019}. Such collisions are also responsible for compaction of these bodies and formation of {a} crust beneath the regolith layer \cite<e.g.,>[]{Bischoff+2006,Blum+2006,Trigo+2006,Rubin2012,Beitz+2016,Tanbakouei+2020}.

{Asteroid families can potentially be used to discuss whether impacts significantly influenced the observed surface materials (Supplementary Table 5). Among ten bodies showing clear $3.1\ {\rm \mu m}$ absorption, seven do not have families, while three (10 Hygiea, 24 Themis, and 128 Nemesis) do. Our spectral analysis suggests ammoniated phyllosilicates for Hygiea, water ice for Themis, and either one or both for Nemesis. Thus, the presence of the ammoniated-phyllosilicate features is mostly consistent with the absence of families which suggests preservation of ammoniated phyllosilicates.}

{Hygiea has been thought to have experienced a disruptive impact \cite{Vernazza+2020}, but it shows an absorption feature consistent with ammoniated phyllosilicates. A possible explanation is incomplete loss of mantle materials upon the impact. Themis likely possesses water ice \cite{Campins2010,Rivkin+Emery2010}, which suggests that the family-forming impact did not lead to complete loss of water ice \cite{Rivkin+2014}. This would also be applicable to Hygiea. Unless the impact completely removed the water-rich mantle, reaccumulation of fragments \cite{Vernazza+2020} and reformation of insulting lag deposit following sublimation of exposed water ice \cite{Schorghofer+2008} would keep the mantle materials.}

{In contrast, if the bodies which do not show the $3.1\ {\rm \mu m}$ absorption have families, it may suggest that impacts are responsible for the lack of ammoniated phillosilicates. An example is 2 Pallas (Supplementary Table 5). Pallas is a heavily-cratered, high density (2.92 $\times$ 10$^3$ ${\rm kg/m^3}$) body \cite{Vernazza+2021}, and thus, it might be possible that Pallas was once water-rock differentiated but lost its water-rich mantle due to the impacts. However, many other bodies without the $3.1\ {\rm \mu m}$ absorption do not have families, suggesting that insufficient water-rock differentiation or the lack or loss of abundant water is a more plausible explanation for the absence of ammoniated phyllosilicates.}

Observational limitation may also cause non-detection. {In} addition to clear detection for {ten} bodies, {two showed} the median $3.1\ {\rm \mu m}$ depths greater than 3\% (but with their 1-$\sigma$ ranges cover 0\%, Supplementary Table 1). We only {included} one D-type asteroid, 361 Bononia, in our {samples} (Figure \ref{fig:features}), but its $3.1\ {\rm \mu m}$ feature also supports a similar outer Solar System origin as suggested from isotopic analysis of carbonates in Tagish Lake meteorites \cite{Fujiya+2019}.

The proposed interior model for large C-complex asteroids can be compared to {that of} Ceres, whose interior structure has been modeled based on gravity and shape measurements \cite{Park+2016,Ermakov+2017,Mao+McKinnon2018}. The inner core of our model is a mineral assemblage with {a} CM/CI-like composition whose density is $\sim$2400--3000 kg/m$^3$ \cite<CM/CI chondrite grain densities,>[]{Consolmagno+2008}. This is an upper limit obtained at W/R = 0.2, and the density can be lower if water ice remains (in the case of W/R$>$0.2). The density of the outer mantle is $\simeq$1200 kg/m$^3$ if W/R = 4 is assumed (20 wt.\% serpentine/saponite mixture {and} 80 wt.\% water ice, {whose densities are} 2500 kg/m$^3$ {and} 900 kg/m$^3$, {respectively}), and it can be lower if W/R$>$4 (more ice remains). This is in good agreement with the two-layer model for Ceres: $\simeq$2400 kg/m$^3$ and 1200--1400 kg/m$^3$ for the inner and outer layers \cite{Ermakov+2017}.

\subsection{Formation and evolution scenario for CCs} 

If ammoniated phyllosilicates are more widespread on outer main belt asteroids than previously thought (Section \ref{subsec:results_AKARI}), their absence in CCs requires {an} explanation. A parsimonious explanation is that CCs and C-complexes formed at similar locations beyond the NH$_3$ and CO$_2$ snow lines at different W/R. While asteroid observations inform the relics of near-surface water-rock reactions, we propose that meteorites chiefly sample the {inner}, low{-}W/R part (Figure \ref{fig:scenario}, Stage 4). {In addition to the fact} that our spectral model for Case 2 with W/R $<$ 1 {reproduced} CCs' spectral features (Section \ref{subsec:results_spectra}), the model mineral composition (Figure \ref{fig:mineral_abundance}) {showed} general agreement with secondary minerals in CM/CI chondrites, namely, the abundant hydrous phyllosilicates (serpentine), organic compounds (represented by pyrene in our geochemical model), pyrrhotite (troilite), magnetite, and carbonates \cite<e.g.,>[]{Rubin1997a,Rubin1997b}. Preferential sampling of rock-dominated cores as CCs is possible because the inner core dominates the volume of an asteroid's rocky material and is less fragile upon collisional disruption compared to the ice-dominated mantle and the upper-most regolith layer, due to lithification caused by secondary mineral formation in pore space \cite<e.g.,>[]{Rubin+2007}. Aqueous alteration takes place also in the water-dominated mantle. However, its high W/R ($>$4) causes water to be mostly left as ice, inter-mixed with rocky grains. Such ice is stable in the subsurface of large asteroids, but {collisionally-produced ice-cemented fragments would be disaggregated as ice is not stable in the inner Solar System. Ice sublimation will} cause efficient disruption of {the fragments} down to the rock grain scale, as we observe main belt comets which release dust likely due to ice sublimation \cite<e.g.,>[]{Hsieh+2006,Hsieh+2018}.

According to our hydrological simulations, W/R of the rocky core would remain low ($<1$) {except for the uppermost grids of the core (the thickness = 2 km)} due to inhibition of effective porewater convection if the interior temperature {was} $<200^\circ$C and/or the planetesimal radius {was} $<500$ km (Section \ref{subsec:results_hydrology}). The former and the latter are applicable for Ceres and the rest of our spectral samples, respectively.
The conditions of low W/R {resulted} in the absence of NH$_4$-bearing saponite and the limited amount of carbonates (Section \ref{subsec:results_geochemical}).
Exhaustion of water from the interior before freezing, if it happened, {would result} in the occurrence of secondary minerals (e.g., chlorite) that are stable at temperature higher than 0${\rm ^\circ C}$ as suggested for some classes of CCs \cite{Zolensky+1993}. 

Petrographic evidence and aggregate accretion theory are in favor of CC parent bodies being water-rock differentiated \cite{Bland+Travis2017}. Furthermore, low W/R independently estimated from oxygen isotopes \cite{Marrocchi+2018} and bulk chemistry \cite{Alexander2019} are particularly consistent with the inner-core origin, otherwise W/R would be comparable or higher than the nebular value ($\sim 1$).
The inner-core origin has been suggested for CI chondrites and Tagish Lake from the lack of their asteroid analogues \cite{Vernazza+2017}, but here we extend the idea to all classes of hydrous CCs. 
{Although} we do not rule out formation inside {the} NH$_3$ and CO$_2$ snow lines for some of {the} CCs (Case 1, Section \ref{subsec:results_geochemical}), bulk elemental and isotopic studies concluded that CC water could have a common origin \cite<e.g.,>[and references therein]{Alexander2019}, {suggesting} that all CCs possibly formed beyond the NH$_3$ and CO$_2$ snow lines.

Organic molecules have been suggested as a possible alternative source of NH$_3$ on Ceres \cite{McSween+2018}, but our {estimation showed} that efficient release of NH$_3$ from IOM {was} required to produce the amount of ammoniated phyllosilicates to cause the observed $3.1\ {\rm \mu m}$ absorption features. 
The highest nitrogen abundance in CC IOM is 2.7 wt.\% \cite{Pizzarello+Williams2012}. Even if the original chondritic rock is assumed to contain 4 wt.\% of IOM, which is the highest organic content in CCs \cite{Alexander+2012}, the nitrogen content in CCs is calculated to be 0.108 wt.\%. This value corresponds to approximately 3.7 wt.\% ammoniated saponite in CCs even if the heaviest member of the smectite solid solution, iron-rich ammoniated saponite, is assumed. 
In comparison, our spectral calculations {indicated} that $>$10 wt.\% of ammoniated saponite {was} required to explain the observed $3.1\ {\rm \mu m}$ {absorption} features (Figures \ref{fig:features} and \ref{fig:mineral_abundance}), though an extensive parameter survey changing grain sizes of all phases {showed} that a lower abundance down to 1 wt.\% {was} possible at least for Ceres \cite{Kurokawa+2020}. 
Moreover, such efficient release of ammonia requires $\sim 300^\circ$C \cite{Pizzarello+Williams2012,McSween+2018}, which is much higher than estimated temperatures experienced by CM/CIs \cite{Krot+2006}. Therefore, we suggest that accretion of NH$_3$ ice provides a better scenario to account for both asteroid observations and analysis of CCs.
We note that icy planetesimals might possess variable abundances of organic compounds. A higher abundance of organics has been proposed for bulk Ceres \cite{Zolotov2020} as its surface likely contains a higher amount of organic compounds than CCs \cite{Prettyman+2017,Kurokawa+2020}.

Assuming that CCs are not fundamentally different from the large C-complex asteroids with $3.1\ {\rm \mu m}$ {absorption} features that occupy much of the mass of the outer main belt, this model explains why the mineralogy of CCs -- the lack of NH$_4$-saponite and the low abundance of carbonates -- differs from the asteroid observations. Telescopic observations of C-complex asteroids and laboratory analysis of CCs are looking at the products in the water-rich, near-surface mantle or rocky core of the same system, respectively. 

{We note that our hypothesis indicating preferential sampling of rock-dominated cores of water-rock differentiated bodies as CCs does not necessarily mean that all CCs originated from the cores. None of Ch and Cgh asteroids in our samples showed $3.1\ {\rm \mu m}$ absorption (Supplementary Table 5). These types of asteroids have been proposed to be the origins of CM2 chondtites \cite{Burbine1998}.}

\subsection{Relevant timescales}

{Our scenario assumed that aqueous alteration proceeded after water-rock differentiation. The timescale of differentiation after ice melting is $\sim$10$^{4-5}$ years \cite{Bland+Travis2017}. The temperature is buffered at 0$^\circ$C for a longer period owing to the latent heat of ice melting \cite<$\sim$10$^{5-6}$ years,>[]{Bland+Travis2017}. Experimental estimates of reaction times at low temperatures are extremely difficult \cite<e.g.,>[]{Jones+Brearley2006}, but the alteration would be very slow. For example, \citeA{Jones+Brearley2006} measured phyllosilicate formation timescales to be days to weeks at 150--200$^\circ$C via experiments using Allende CV3 carbonaceous chondrite, and estimated the alteration timescale to be $\sim$10$^{4\pm2}$ years at 25$^\circ$C by extrapolation. It would be much slower at 0°C. Additionally, extrapolating dissolution rates down to 0$^\circ$C led to the timescale estimation to be $\sim$10$^5$ years and $\sim$10$^6$ years for 100 ${\rm \mu m}$ grains of olivine and enstatite, respectively, at pH $\sim$12 \cite<estimated based on kinetic parameters given by>[]{Palandri+Kharaka2004}. Thus, although some reactions might have proceeded during sedimentation of grains, the majority of aqueous alterations would have proceeded after water-rock differentiation. }

{Once internal heating overcomes the latent heat of ice melting, the temperature increases above 0$^\circ$C.} The water-rock reactions assumed in our model {are} faster than the fluid circulation (Myrs) in the hypothetical planetesimals (200 and 500 km in radius) {at, for instance, 25$^\circ$C \cite{Jones+Brearley2006}}, suggesting that a near-equilibrium state between rock and fluid could be maintained each time and everywhere in planetesimals even if W/R and temperature slowly changed during global fluid circulation (of course, fluid vaporization during water-rock reactions and textural insulation by alteration minerals can generate large- and small-scale disequilibria, respectively, as inferred from meteorites). The sluggish flow may explain CC features {which suggest} the dominance of static aqueous alteration \cite{Trigo+Rubin2006,Rubin+2007,Trigo+2019}, while fluid flow in planetesimals \cite{Young+1999} was assumed in our model. 

In the proposed scenario (Figure \ref{fig:scenario}), accretion (Stage 1) should have {been} completed {at} $<$4 Myrs from calcium–aluminium-rich inclusion formation to initiate water-rock differentiation \cite<e.g.,>[]{Bland+Travis2017}. This timescale has {also} been constrained from the onset time of aqueous alteration informed from {the} Mn-Cr system in CC carbonates \cite{Fujiya+2013,Lee+2012,Lee+2013}. The accretion timescale is consistent with dust coagulation models \cite<e.g.,>[]{Okuzumi+2012} and with the existence of substructures in Myrs-old protoplanetary disks in a few tens of au scale \cite<e.g.,>[]{Andrews+2018}, which may be signatures of ongoing planetesimal formation. After accretion, differentiation and alteration started in $\sim$1 Myr \cite<e.g.,>[]{Bland+Travis2017}. Time of freezing (Stage 3) depends on the sizes of planetesimals \cite{Wakita+Sekiya2011}. Complete H$_2$O loss via sublimation to space depends on the size and heliospheric distance \cite{Schorghofer+2008} and likely takes longer than the age of the Solar System for large asteroids, as water ice likely exists in Ceres and on Themis \cite{Campins2010,Rivkin+Emery2010}. We note that the proposed scenario for the distant origin is not sensitive to the timing of migration, {as} long as the majority of accretion takes place prior to migration.

\subsection{Implications for planet formation} 

Our hypothesis that C-complex asteroids and CC parent bodies accreted NH$_3$ and CO$_2$ ice is new, but it is naturally expected from the modern planet formation theory. Radial mixing of small bodies due to gravitational scattering by gas giants in the Grand Tack scenario \cite<e.g.,>[]{Walsh+2011} assumes that C-type asteroids originate from near or beyond giant planet formation {regions}, namely, beyond {the} NH$_3$ and CO$_2$ snow lines (currently $>10$ au). Alternatively, NH$_3$ and CO$_2$ ice might be delivered directly to bodies formed in the main belt via icy pebbles, following the inward migration of these snow lines in the late stage of the solar nebula evolution \cite{DeSanctis+2015,Nara+2019}. Direct evidence of icy pebble accretion is difficult to find, but a carbon-rich clast in a CR chondrite \cite{Nittler+2019} might be an example of migration and accretion of such small fragments. 

We {proposed} that {C-complex} and CCs parent bodies formed with a comet-like composition, but this does not necessarily mean that C-complex asteroid precursors are equivalent to comets. Rather, the difference in deuterium to hydrogen ratio between CCs and comets \cite{Alexander+2012} suggest that they formed from different reservoirs in the solar nebula. In addition to the NC/CC dichotomy \cite{Kruijer+2017}, there should {have been} at least another distinct reservoir to separate CC- and comet-building materials in the early Solar System. Our study {suggests} that at least the latter boundary was located beyond {the} NH$_3$ and CO$_2$ snow lines. Thus, mechanisms to partition reservoirs in the solar nebula were not limited to early formation of Jupiter {\cite{Kruijer+2017,Desch+2018}}. A pressure maximum near Jupiter’s location in the solar nebula, which halts inward migration of pebbles, has been proposed as an alternative origin of NC/CC dichotomy \cite{Brasser+Mojzsis2020}. We suggest that there might be multiple pressure {traps} to separate reservoirs. {Possible origins of pressure traps include gas-dust viscous gravitational instability \cite{Tominaga+2019}, magneto-hydrodynamical wind \cite{Taki+2021}, sublimation near the snow line(s) \cite{Charnoz+2021}, and disk–planet interaction \cite{Kanagawa+2015}.} Such multiple pressure {traps} in a single protoplanetary disk are thought to create ringed structures which were found to be common in protoplanetary disks observed recently by the Atacama Large Millimeter/submillimeter Array \cite<the spacial scales of rings are several tens of au,>[]{Andrews+2018,Dullemond+2018}. A ringed structure supported by multiple pressure {traps} may be important for formation of planetesimals and planets in our Solar System, as well as extrasolar systems \cite{Johansen+2014,Morbidelli+2020}.

Bulk W/R of CC parent bodies {was} proposed to be higher than those recorded {for} CCs in our study, {which} may explain the disagreement in volatile compositions between bulk silicate Earth (BSE) and chondrites. BSE has higher hydrogen to carbon ratio (H/C) than {the} chondritic value \cite{Hirschmann+Dasgupta2009,Dasgupta+Grewal2019}, which {was} expected in our model. A fraction of small bodies transferred inward from the outer Solar System would have been captured in the main belt, while others accreted onto Earth during its formation, contributing substantial water and leading to Earth's high H/C ratio.

\subsection{Implications for sample return missions and future observations} 
\label{subsec:discussion_future}

The proposed new paradigm can be tested in {the} near future by return samples from small carbonaceous asteroids by Hayabusa2 \cite{Watanabe+2019} and OSIRIS-REx \cite{Lauretta2+017}. Our model (Case 2, W/R $<1$, Section \ref{subsec:results_geochemical}) predicts that returned samples may contain trace amounts of NH$_4$-bearing salts (halite) and/or NH$_3$- (and possibly CO$_2$-) bearing fluid inclusions. These would be easily lost from CCs in terrestrial environments due to deliquescence, as they are rarely found in meteorites except for few examples, such as halite in Zag and Monahans (1998) \cite{Zolensky+1999,Whitby+2000,Rubin+2002,Chan+2018}, but could be found in returned samples. Remote detection of carbonates \cite{Kaplan+2020} and possible existence of nanophase magnetites which causes its \textquotedblleft blue\textquotedblright \ spectral slope due to space weathering on Bennu \cite{Trang+2021} are also consistent with model prediction (a few wt.\% in Case 2 at W/R $<1$ and $T=0^\circ$C, Figure \ref{fig:mineral_abundance}). 

{Large C-complex main-belt asteroids with ammoniated-phyllosilicate features were proposed to have the same origin with CCs' parent bodies in our scenario. The sample return from these asteroids \cite<e.g., from Ceres,>[]{Burbine+Greenwood2020,Gassot+2021,Castillo2021} is ultimately needed to understand their building blocks and evolution history. The proposed scenario predicts that they would show the isotopic similarity to CCs, but not to comets.} After the submission of this study, \citeA{Tsuchiyama+2021} reported CO$_2$ fluid trapped in a CC. They suggested that its parent body accreted CO$_2$ ice beyond {the} CO$_2$ snow line, which is fully consistent with our scenario for {the} CC {origins}.

{Future asteroid observations are also useful to test the proposed model.} Our geochemical model {predicts} that the abundance of carbonates correlates with that of ammoniated saponite (Section \ref{subsec:results_geochemical}), which should be visible in their $3.4$ and $4.0\ {\rm \mu m}$ absorptions (Section \ref{subsec:results_spectra}). Carbonates were found on Ceres \cite{DeSanctis+2015} as well as Bennu \cite{Kaplan+2020}. AKARI and IRTF spectra {at} these wavelengths do not agree with each other, suggesting {that} they are severely influenced by, for example, the thermal component (Section \ref{subsec:model_data}). Observations of asteroids showing $3.1\ {\rm \mu m}$ absorption with improved accuracy {at} these longer wavelengths would be particularly useful to test our hypothesis. {Moreover, increasing the number of samples, especially including smaller sizes are important. Although our samples are mostly larger than 100 km and do not show clear correlation between the size and ammoniated-phyllosilicate features (Supplementary Figure 6), we predict that much smaller bodies tend to lack these features. }

\section{Conclusions}
\label{sec:conclusions}

We combined spectral analysis of large outer main belt asteroids observed {using the} AKARI space telescope ($>2.5\ {\rm \mu m}$) and hydrological, geochemical, spectral modeling of water-rock interactions in icy planetesimals and their products to understand the origin of C-complex asteroids showing $3.1\ {\rm \mu m}$ absorption features and their relations to CCs. {We have come to the following conclusions.} i) Our spectral analysis showed that the positions and widths of the absorption band are more consistent with ammoniated phyllosilicates than water ice for the majority of samples. Brucite was also ruled out as the source of absorption features by combining ground-based data for shorter wavelengths ($<2.5\ {\rm \mu m}$). ii) Our geochemical model combined {with} synthetic spectral calculations showed that the surface materials of outer main belt asteroids showing $3.1\ {\rm \mu m}$ absorption features {of} ammoniated phyllosilicates can form from starting materials containing NH$_3$ and CO$_2$ ice {through} water-rock reactions in water-rich (W/R $>4$) and low temperature ($<70^\circ$C) conditions. In contrast, CC-like mineral assemblages forms in relatively water-poor (W/R $<4$) and/or high temperature ($>70^\circ$C) conditions. iii) Our hydrological model demonstrated that such difference in W/R can be maintained in a single, water-rock differentiated body, if the interior temperature is $\lesssim 100^\circ$C (for the largest asteroid Ceres) and/or the planetesimal radius is $<500$ km (all of our spectral samples except for Ceres). 

Based on these results, we proposed that several, {if not} all, C-complex asteroids and CC parent bodies formed beyond NH$_3$ and CO$_2$ snow lines and differentiated. Surface minerals containing ammoniated phyllosilicates and CC-like materials formed in the water-rich mantles and in the rock-dominated cores of the water-rock differentiated bodies, respectively. The materials in the rocky cores are less fragile due to lithification and thus more likely to be sampled as meteorites. The distant origin of C-complex asteroids is naturally expected from the modern planet formation theory which involves Solar System-scale migration of pebbles and planetesimals. Moreover, the hypothesized water-rich progenitors of C-complex asteroids and CCs can, if they accreted onto early Earth, explain {why the H/C ration of} Earth {is higher than that of} CCs. Finally, our hypothesis can be {tested via} sample return missions by looking for NH$_4$-bearing salts (halite) and/or NH$_3$-(and possibly CO$_2$-) bearing fluid inclusions, {as well as} future asteroid observations by testing the correlation of $3.1\ {\rm \mu m}$ ammoniated phase {with} $3.4\ {\rm \mu m}$ and $4.0\ {\rm \mu m}$ carbonate features.

\section*{Author contributions statement}

H.K. and B.E. performed the spectral calculations. 
T.S. and S.K. performed the thermodynamic calculations for water-rock reactions.
Y.S. performed water circulation simulations. 
F.U. compiled the asteroid observation data. 
M.Y. performed thermodynamic calculations for ice mixtures. 
All authors designed the project and wrote the manuscript.

\section*{Additional information}

The authors declare no competing interests.

\newpage

\begin{table}[p]
    \centering
    \begin{tabular}{lrrrl}
        Name & CO$_2$ [mol\%] & H$_2$S [mol\%] & NH$_3$ [mol\%] & Note \\ \hline
        Case 1 & 0 & 0 & 0 & Pure water and rock \\
        Case 2 & 1 & 0.5 & 0.5 & Cometary composition, CO$_2$ depleted \\
        Case 3 & 10 & 0.5 & 0.5 & Cometary composition \\
    \end{tabular}
    \caption{Volatile abundance relative to water in our geochemical model. 
    }
    \label{tab:model}
\end{table}

\newpage

\begin{figure}[p]
    \centering
    \includegraphics[width=1.0\linewidth]{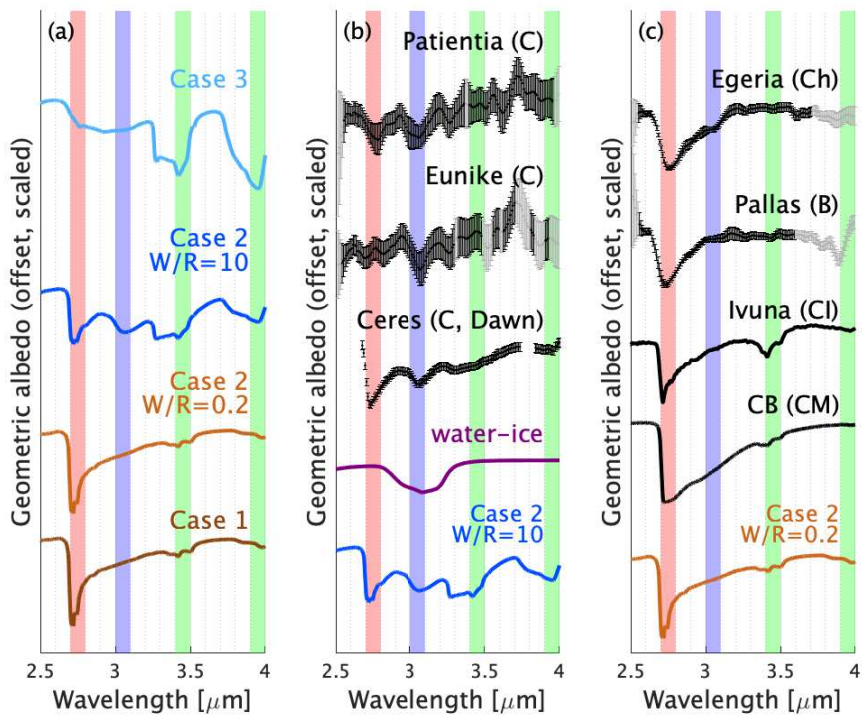}
    \caption{Comparison between models and observed infrared reflectance spectra. (a) Model spectra. Brown line: Case 1, $T$ = 0$^\circ$C, and W/R = 10. Orange line: Case 2, $T$ = 0$^\circ$C, and W/R = 0.2. Blue line: Case 2, $T$ = 0$^\circ$C, and W/R = 10. Cyan line: Case 3, $T$ = 0$^\circ$C, and W/R = 10. (b) Spectra of asteroids showing $2.7$, $3.1$, and $3.4\ {\rm \mu m}$ absorption features compared to the Case 2 model and the water-ice coating model \cite{Rivkin+Emery2010}. (c) Spectra of asteroids and CM/CI chondrites showing a dominant $2.7\ {\rm \mu m}$ absorption feature. Observed spectra were scaled and offset to compare the absorption features (see Supplementary Figures 1 and 10 for the original spectra). The gray data points indicate unreliable wavelength regions due to large uncertainties defined by \citeA{Usui+2019}. We highlighted the position of three absorption features. Red {area}: $2.7\ {\rm \mu m}$ (hydrous minerals). Blue {area}: $3.1\ {\rm \mu m}$ (ammoniated phyllosilicates or water ice). Green {area}: $3.4$ and $4.0\ {\rm \mu m}$ (carbonates). AKARI, Dawn, and meteorite data are from \citeA{Usui+2019}, \citeA{Ciarniello+2017}, and the references in Supplementary Table 4, respectively.}
    \label{fig:spectra}
\end{figure}

\newpage

\begin{figure}[p]
    \centering
    \includegraphics[width=1.0\linewidth]{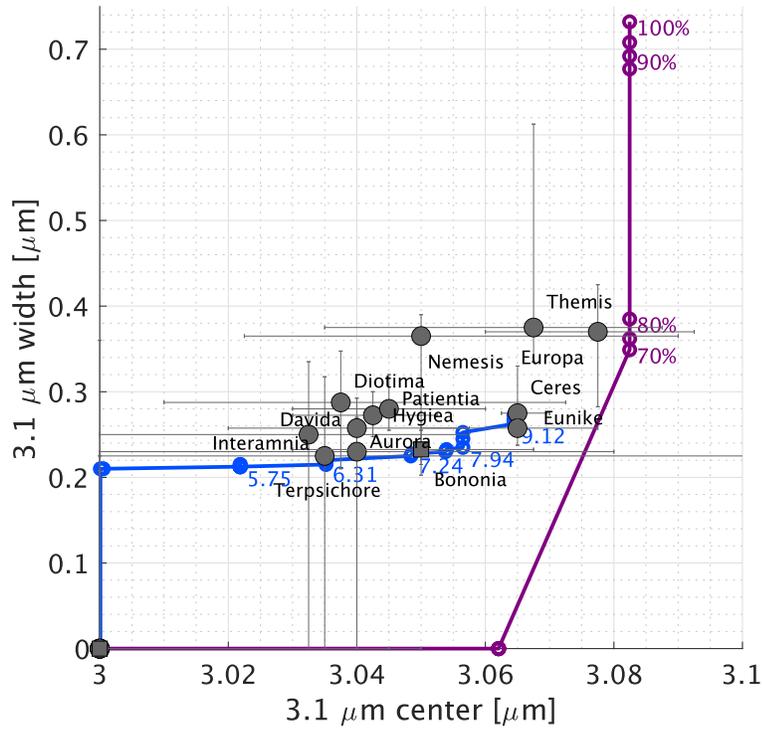}
    \caption{Center and width of the $3.1\ {\rm \mu m}$ absorptions for asteroids observed {using} AKARI (gray circles and squares for C-complex and D- and T-types, respectively) compared to a successive set of mineral assemblages from Case 2 {alterations} with changing W/R (blue line with W/R values given; 0$^\circ$C model) and, alternatively, linear mixing fractions of the water-ice coating model \cite{Rivkin+Emery2010} with an observed asteroid (2 Pallas) that lacks $3.1\ {\rm \mu m}$ absorption (purple line with the values of water-ice coating model fraction given). The 1-$\sigma$ dispersions are presented by error bars. The values are listed in Supplementary Table 1.}
    \label{fig:31um}
\end{figure}

\newpage

\begin{figure}[p]
    \centering
    \includegraphics[width=1.0\linewidth]{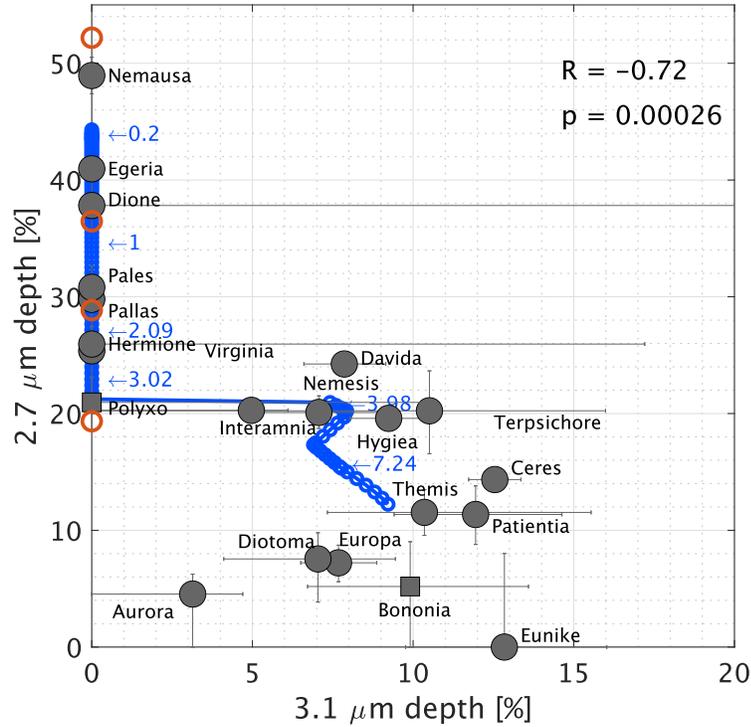}
    \caption{Comparison of absorption features between the model and observations. Data points correspond to C-complex and D- and T-type asteroids observed {using} AKARI (filled gray circles and squares) and CI and CM carbonaceous chondrites (open orange circles, Supplementary Table 4), respectively. The 1-$\sigma$ dispersions are presented by error bars. The values are listed in Supplementary Table 1. The linear correlation coefficient and its significance level are $R = -0.72$ and $p = 0.00026$, both of which support that the two values are (anti-)correlated. The blue line {represents} Case 2 {at} 0$^\circ$C with {the} changing the initial {W/R} input (the W/R values are given in the figure). Grain sizes are $0.5$ and $100\ {\rm \mu m}$ for IOM and the other phases, respectively (see Supplementary Figure 11, for the grain size {dependence}). }
    \label{fig:features}
\end{figure}

\newpage

\begin{figure}[p]
    \centering
    \includegraphics[width=1.0\linewidth]{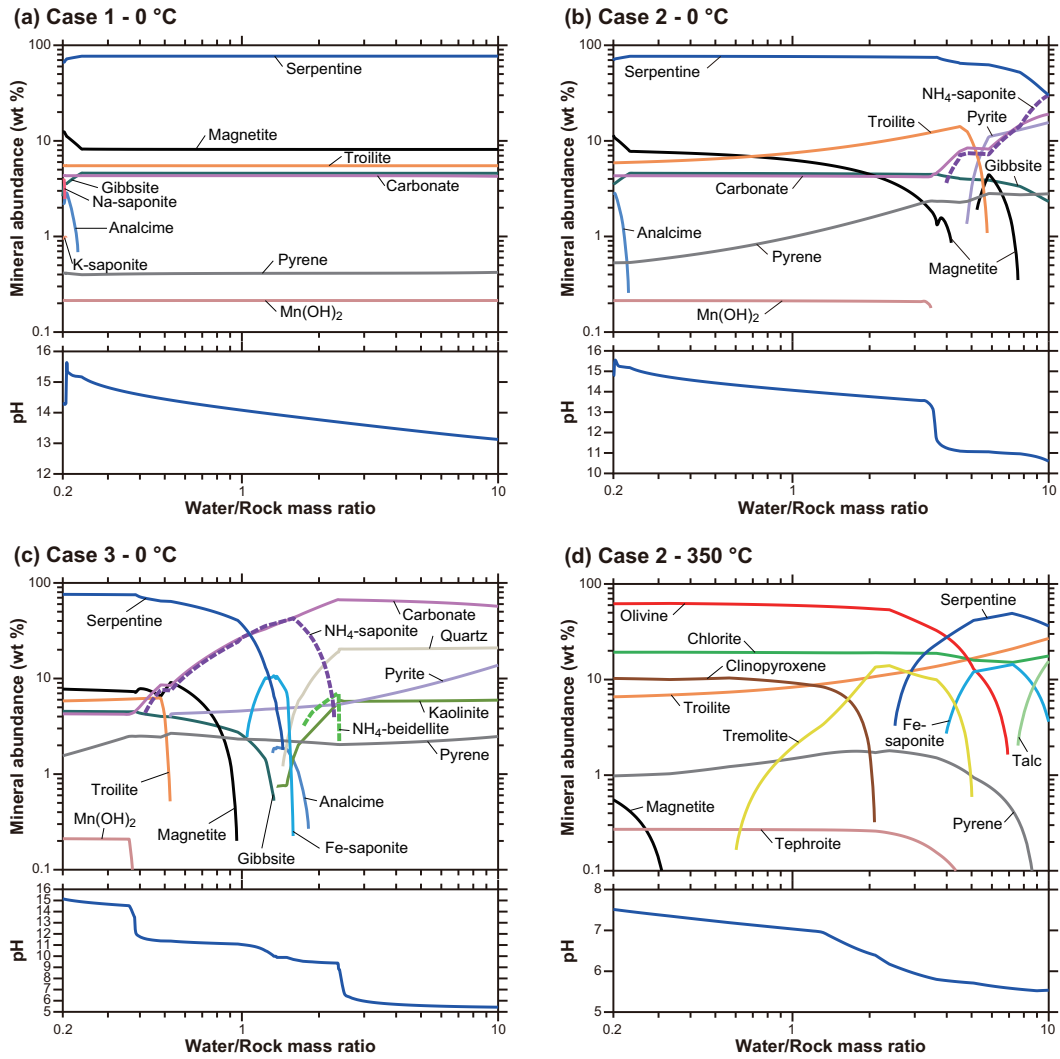}
    \caption{Mineral abundance and pH obtained from thermodynamic modeling as a function of W/R. (a) Case 1, $T = 0^\circ$C. (b) Case 2, $T = 0^\circ$C. (c) Case 3, $T = 0^\circ$C. (d) Case 2, $T = 350^\circ$C.}
    \label{fig:mineral_abundance}
\end{figure}

\newpage

\begin{figure}
    \centering
    \includegraphics[width=0.6\linewidth]{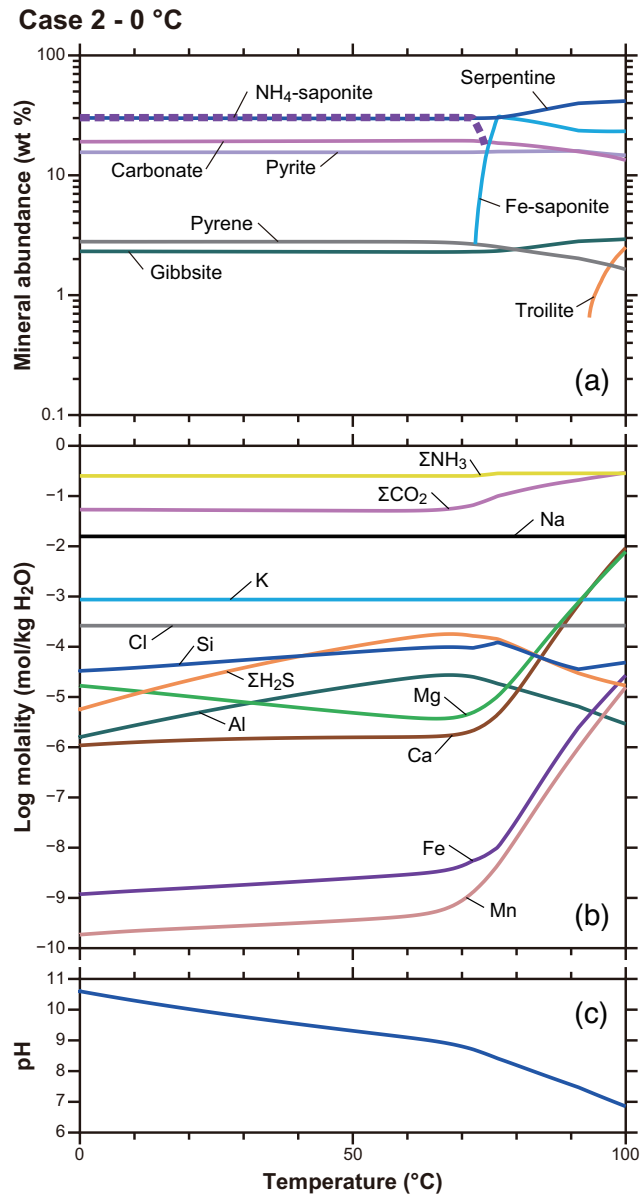}
    \caption{Temperature dependency of (a) abundance of alteration minerals, (b) concentrations of dissolved element and species in fluid, and (c) in-situ pH of fluid at W/R = 10 in Case 2, $T=0^\circ$C.}
    \label{fig:Tdependence}
\end{figure}

\newpage

\begin{landscape}
\begin{figure}[p]
    \centering
    \includegraphics[width=1.0\linewidth]{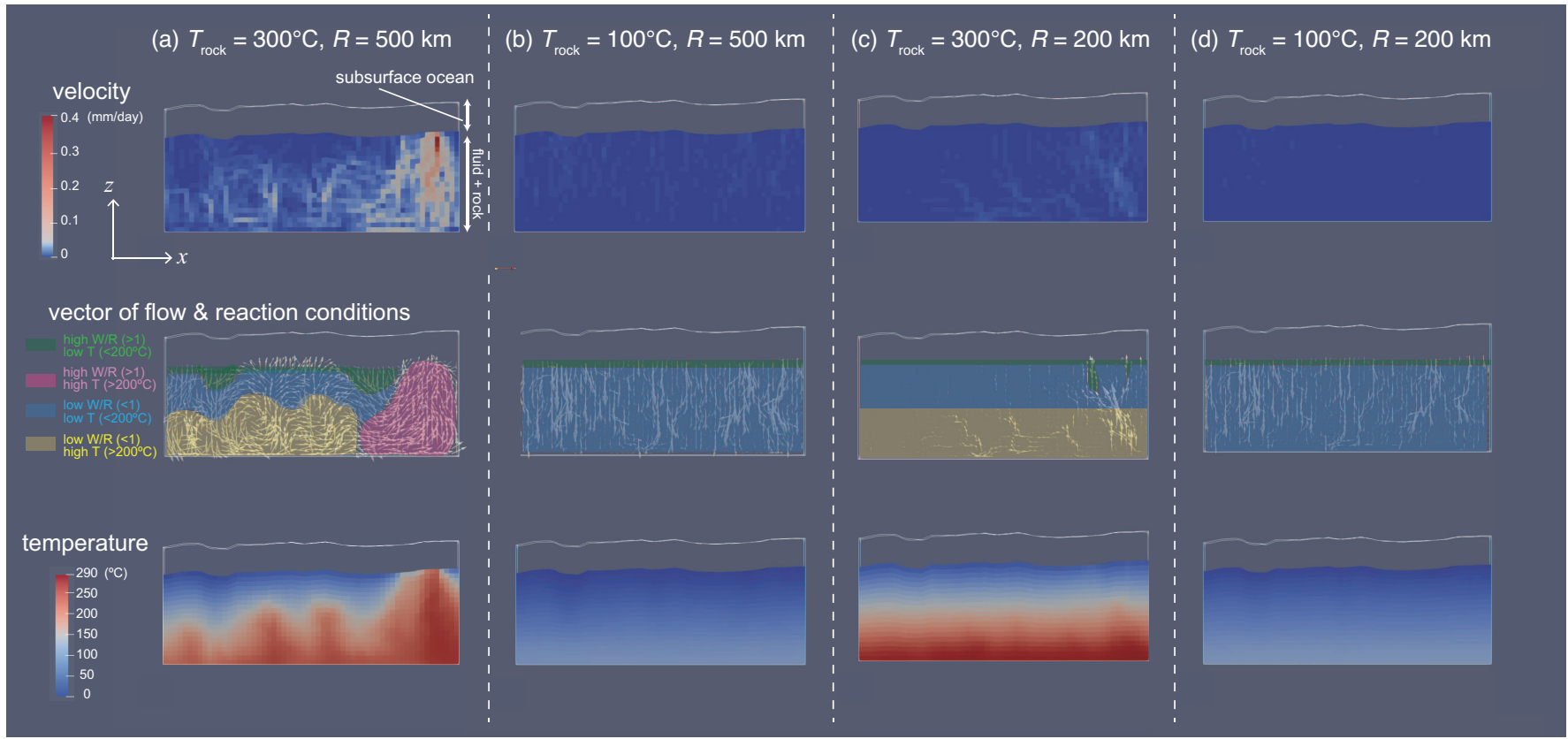}
    \caption{Results of hydrological simulations (snapshot of velocity, vector, and temperature of fluids in each case) for (a) initial core temperature, $T_{\rm rock}$, = 300$^\circ$C and planetesimal radius, $R$, = 500 km, (b) $T_{\rm rock}$ = 100$^\circ$C and $R$ = 500 km, (c) $T_{\rm rock}$ = 300$^\circ$C and $R$ = 200 km, and (d) $T_{\rm rock}$ = 100$^\circ$C and $R$ = 200 km. The white flames represent the calculated area of the two-dimensional grid-block system (total length 240 km and total depth $\sim$50 km). The snapshots {were} taken {at} 5 Myrs for the 200-km planetesimal and 8 Myrs for the 500-km planetesimal after {starting} the simulations. The distribution of approximate W/R is indicated in the middle panels. Characteristic minerals expected to form are: NH$_4$-saponite and carbonates (high W/R and low $T$, green), carbonates (high W/R and high $T$, red), serpentine (low W/R and low $T$, blue), and unhydrous minerals (high W/R and high $T$, yellow). See Fig. \ref{fig:mineral_abundance} for details.}
    \label{fig:water_circulation}
\end{figure}
\end{landscape}

\newpage

\begin{figure}[p]
    \centering
    \includegraphics[width=1.0\linewidth]{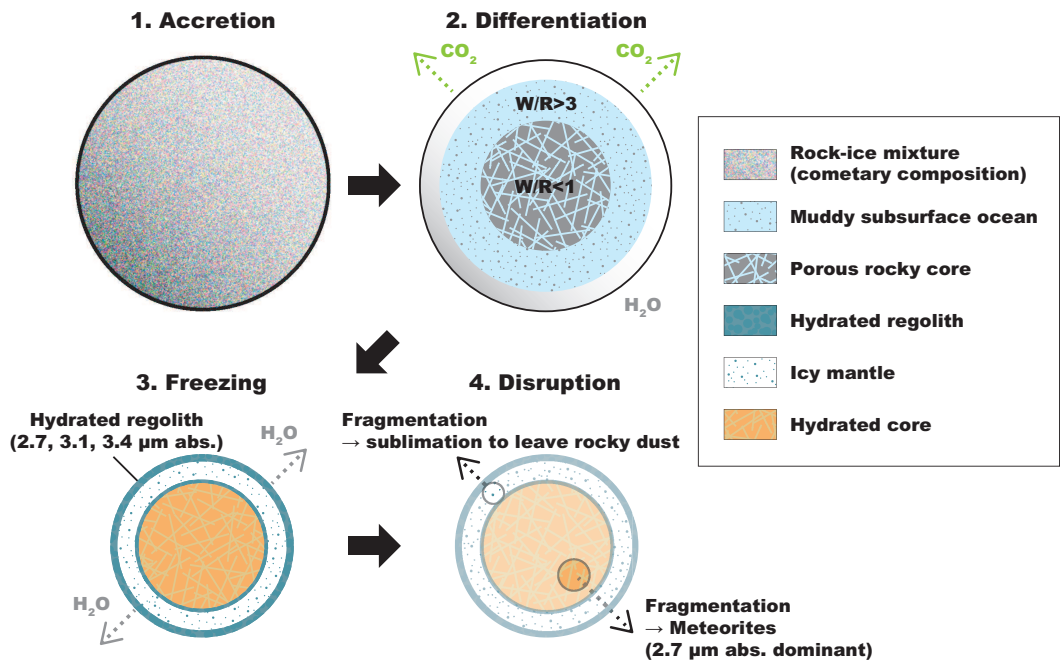}
    \caption{A scenario for the formation and evolution of large C-complex asteroids in the main belt and {its relationship} to carbonaceous chondrites. Stage 1: Accretion. Stage 2: Differentiation and alteration. Stage 3: Freezing ({length} depends on the size). Stage 4: Catastrophic disruption. }
    \label{fig:scenario}
\end{figure}

\newpage

%%%%%%%%%%%%%%%%%%%%%%%%%%%%%%%%
%% Optional Appendix goes here
%
% The \appendix command resets counters and redefines section heads
%
% After typing \appendix
%
%\section{Here Is Appendix Title}
% will show
% A: Here Is Appendix Title
%
%\appendix
%\section{Here is a sample appendix}

\appendix

\section{Text S1. Thermodynamic modeling of the ice mixtures}

In order to estimate the volatile loss from icy planetesimals to space via degassing as a consequence of heating by radioactive decay in Stage 1 in Figure 7, we computed the phase change of the ice mixture in a closed system with FREZCHEM \cite{Marion+Kargel2007,Marion+2012,Zolotov2017}. 
FREZCHEM calculates the chemical equilibrium between dissolved species in the solution, solid phases precipitated from the solution, and gas phases at each temperature and pressure upon freezing with the Gibbs free energy minimization method \cite{mironenko1997frezchem2}. 
As we are interested in gas release upon heating, the solid phases which could form upon freezing (e.g. NH$_4$CO$_3$) were suppressed. 
The equilibrium of gas phases with the solution was controlled by Henry’s law. 
Gas hydrate phases were also considered by extrapolating Henry’s law constants below 0$^\circ$C.
We considered the temperature from 25$^\circ$C to $-$10$^\circ$C. 
A wide pressure range from 1 bar (near the surface) to 1000 bar (the center of a Ceres-sized body) was considered.
We assumed a cometary composition (Case 3, Table 1). We neglected H$_2$S for simplicity.

The abundances of gas, liquid, and solid phases are shown in Supplementary Figure \ref{figS:ice_mixture}.
At 1 bar, 95 mol\% of CO$_2$ was in a gas phase regardless of the dominant phase of H$_2$O. 
A minor amount of CO$_2$ was dissolved in the solution as H$_2$CO$_3$, HCO$_3$$^-$ and CO$_3$$^{2-}$. 
The ratios of HCO$_3$$^-$ and CO$_3$$^{2-}$ to H$_2$CO$_3$ were controlled by charge balance in the solution, and the neutral pH limited the total abundance of the dissolved CO$_2$ phases. 
In contrast, nearly 100 mol\% of NH$_3$ was dissolved in the solution as NH$_4^+$. 
At higher pressure, CO$_2$ gas hydrate and dissolved phases became dominant CO$_2$ reservoirs at lower and higher temperature, respectively. 

These results suggest that, as temperature increases due to radioactive decay heating, CO$_2$ ice would melt and be released as CO$_2$ gas from a near surface layer even before the melting of H$_2$O ice.
Melting of H$_2$O ice and the subsequent initiation of convection would allow the materials which existed in the deep interior where CO$_2$ gas hydrate and dissolved phases are stable to once reach the near surface, which led to the release of CO$_2$ gas.
Though the water-rock reactions potentially raise pH and stabilize the dissolved CO$_2$ phases, the chemical reactions are likely to be slower than the CO$_2$ phase change and the release of CO$_2$ gas.
As a result, the icy planetesimals, especially their outer parts, would become depleted in CO$_2$ compared to the original composition whereas NH$_3$ is retained, modeled as Case 2 (Table 1).

\newpage

%Type or paste text here. This should be additional explanatory text, such as: extended descriptions of results, full details of models, extended lists of acknowledgements etc.  It should not be additional discussion, analysis, interpretation or critique. It should not be an additional scientific experiment or paper.
%
%Repeat for any additional Supporting Text

%%Enter Data Set, Movie, and Audio captions here
%%EXAMPLE CAPTIONS

\begin{table}[p]
    \centering
    \renewcommand{\arraystretch}{1.2}
    \begin{tabular}{lllll}
    Asteroid & $D_{2.7}$ [\%] & $D_{3.1}$ [\%] & $W_{3.1}\ {\rm [\mu m]}$ & $C_{3.1}\ {\rm [\mu m]}$  \\  \hline
    	0001 Ceres (AKARI) & 14.3$_{-0.8}^{+0.8}$ & 12.5$_{-0.8}^{+0.8}$ & 0.275$_{-0.008}^{+0.008}$ & 3.065$_{-0.003}^{+0.005}$ \\
0001 Ceres (IRTF) & -- & 9.35$_{-0.16}^{+0.16}$ & 0.304$_{-0.004}^{+0.004}$ & 3.047$_{-0}^{+0.004}$ \\
0001 Ceres (Dawn) & 27.3$_{-1.4}^{+1.5}$ & 16.0$_{-4.8}^{+1.7}$ & 0.283$_{-0.015}^{+0.018}$ & 3.065$_{-0.008}^{+0.008}$ \\
0002 Pallas (AKARI) & 29.8$_{-0.9}^{+0.9}$ & 0$_{-0}^{+0}$ & 0$_{-0}^{+0}$ & -- \\
0002 Pallas (IRTF) & -- & 5.09$_{-0.42}^{+0.40}$ & 0.33$_{-0.02}^{+0.06}$ & 3.041$_{-0.005}^{+0.005}$ \\
0010 Hygiea (AKARI) & 19.6$_{-0.8}^{+0.8}$ & 9.24$_{-1.19}^{+1.31}$ & 0.272$_{-0.025}^{+0.028}$ & 3.042$_{-0.013}^{+0.010}$ \\
0013 Egeria (AKARI) & 41.0$_{-0.9}^{+0.9}$ & 0$_{-0}^{+0}$ & 0$_{-0}^{+0}$ & -- \\
0013 Egeria (IRTF) & -- & 0$_{-0}^{+0}$ & 0$_{-0}^{+0}$ & -- \\
0024 Themis (AKARI) & 11.5$_{-2.0}^{+1.9}$ & 10.4$_{-3.0}^{+5.2}$ & 0.375$_{-0.040}^{+0.237}$ & 3.067$_{-0.033}^{+0.020}$ \\
0024 Themis (IRTF) & -- & 7.85$_{-1.19}^{+1.15}$ & 0.423$_{-0.060}^{+0.224}$ & 3.096$_{-0.014}^{+0.011}$ \\
0049 Pales (AKARI) & 30.8$_{-1.9}^{+1.8}$ & 0$_{-0}^{+0}$ & 0$_{-0}^{+0}$ & -- \\
0050 Virginia (AKARI) & 25.4$_{-5.0}^{+3.4}$ & 0$_{-0}^{+0}$ & 0$_{-0}^{+0}$ & -- \\
0051 Nemausa (AKARI) & 48.9$_{-1.6}^{+1.6}$ & 0$_{-0}^{+0}$ & 0$_{-0}^{+0}$ & -- \\
0051 Nemausa (IRTF) & -- & 0$_{-0}^{+0}$ & 0$_{-0}^{+0}$ & -- \\
0052 Europa (AKARI) & 7.22$_{-1.62}^{+1.51} $ & 7.68$_{-1.17}^{+1.19}$ & 0.370$_{-0.088}^{+0.055}$ & 3.078$_{-0.018}^{+0.015}$ \\
0052 Europa (IRTF) & -- & 4.96$_{-3.05}^{+0.68}$ & 0.63$_{-0.34}^{+0.03}$ & 3.036$_{-0.010}^{+0.017}$ \\
0081 Terpsichore (AKARI) & 20.2$_{-3.7}^{+3.4}$ & 10.5$_{-10.5}^{+5.5}$ & 0.225$_{-0.225}^{+0.093}$ & 3.035$_{-0.035}^{+0.123}$ \\
0094 Aurora (AKARI) & 4.54$_{-4.54}^{+1.70}$ & 3.14$_{-3.14}^{+1.56}$ & 0.230$_{-0.230}^{+0.040}$ & 3.040$_{-0.040}^{+0.040}$ \\
0106 Dione (AKARI) & 37.8$_{-2.1}^{+2.2}$ & 0$_{-0}^{+20.9}$ & 0$_{-0}^{+0.208}$ & 3$_{-0}^{+0.050}$ \\
0121 Hermione (AKARI) & 25.9$_{-1.5}^{+1.5}$ & 0$_{-0}^{+17.2}$ & 0$_{-0}^{+0.360}$ & 3.000$_{-0}^{+0.040}$ \\
0121 Hermione (IRTF) & -- & 0$_{-0}^{+0}$ & 0$_{-0}^{+0}$ & -- \\
0128 Nemesis (AKARI) & 20.1$_{-1.4}^{+1.4}$ & 7.07$_{-2.10}^{+2.03}$ & 0.365$_{-0.163}^{+0.025}$ & 3.050$_{-0.028}^{+0.040}$ \\
0185 Eunike (AKARI) & 0$_{-0}^{+8.02}$ & 12.8$_{-3.1}^{+3.2}$ & 0.257$_{-0.020}^{+0.073}$ & 3.065$_{-0.015}^{+0.013}$ \\
0185 Eunike (IRTF) & -- & 3.35$_{-3.35}^{+3.31}$ & 0.193$_{-0.193}^{+0.206}$ & 3.015$_{-0.016}^{+0.067}$ \\
0308 Polyxo (AKARI) & 21.0$_{-2.4}^{+2.1}$ & 0$_{-0}^{+10.5}$ & 0$_{-0}^{+0.225}$ & 3.000$_{-0}^{+0.045}$ \\
0308 Polyxo (IRTF) & -- & 0$_{-0}^{+0}$ & 0$_{-0}^{+0}$ & -- \\
0361 Bononia (AKARI) & 5.19$_{-5.19}^{+3.84}$ & 9.91$_{-3.19}^{+3.68}$ & 0.232$_{-0.018}^{+0.023}$ & 3.05$_{-0.02}^{+0.018}$ \\
0361 Bononia (IRTF) & -- & 1.79$_{-1.79}^{+0.89}$ & 0.284$_{-0.284}^{+0.070}$ & 3.033$_{-0.033}^{+0.021}$ \\
0423 Diotima (AKARI) & 7.55$_{-3.70}^{+2.25}$ & 7.04$_{-2.93}^{+2.41}$ & 0.288$_{-0.078}^{+0.060}$ & 3.038$_{-0.028}^{+0.035}$ \\
0451 Patientia (AKARI) & 11.4$_{-2.6}^{+2.5}$ & 11.9$_{-2.5}^{+2.7}$ & 0.28$_{-0.03}^{+0.04}$ & 3.045$_{-0.015}^{+0.015}$ \\
0451 Patientia (IRTF) & -- & 5.04$_{-1.95}^{+1.96}$ & 0.281$_{-0.077}^{+0.091}$ & 3.033$_{-0.013}^{+0.031}$ \\
0511 Davida (AKARI) & 24.2$_{-0.9}^{+1.0}$ & 7.86$_{-1.26}^{+1.29}$ & 0.258$_{-0.020}^{+0.035}$ & 3.040$_{-0.020}^{+0.0175}$ \\
0511 Davida (IRTF) & -- & 0$_{-0}^{+0}$ & 0$_{-0}^{+0}$ & -- \\
0704 Interamnia (AKARI) & 20.3$_{-0.9}^{+0.9}$ & 4.97$_{-4.97}^{+1.14}$ & 0.25$_{-0.25}^{+0.09}$ & 3.033$_{-0.033}^{+0.025}$ \\
0704 Interamnia (IRTF) & -- & 2.73$_{-1.07}^{+1.18}$ & 0.302$_{-0.178}^{+0.082}$ & 3.030$_{-0.011}^{+0.013}$ \\

    \end{tabular}
    \caption{$2.7\ {\rm \mu m}$ absorption depth ($D_{2.7}$), $3.1\ {\rm \mu m}$ absorption depth ($D_{3.1}$), width ($W_{3.1}$), and center ($C_{3.1}$) of asteroids spectra acquired using AKARI, IRTF, and Dawn. Median and 68\% confidence intervals are shown.}
    \label{tabS:absorption_depth}
\end{table}

\newpage

\begin{table}[p]
    \centering
\begin{tabular}{lll}
    Endmember & RELAB ID or reference & n \\
    \hline
    Alabandite & WV-JWH-005 (BKR1WV005) & 2.73 \\ %ok
    Analcime & USGS, GDS1 & 1.495 \\ %ok
    NH$_4$-saponite & Ehlmann et al.\cite{Ehlmann+2018} & 1.56 \\ %ok
    %NH$_4$-beidellite & BE-EAC-003 (LABE03) & 1.46 & Beidellite \\
    Andradite & GN-EAC-004 (LAGN04) & 1.77 \\ %ok
    Beidellite & BE-EAC-003 (LABE03) & 1.46 \\ %ok
    Brucite & JB-JLB-944-A (BKR1JB944A) & 1.58 \\ %ok
    Calcite & CB-EAC-010-A (LACB10A) & 1.57 \\ %ok
    Chrysotile & CR-TXH-006 (LACR06) & 1.57 \\ %ok
    Clinochlore & CL-TXH-014 (LACL14) & 1.59 \\ %ok
    Diaspore & HO-EAC-007-A (LAHO07A) & 1.495 \\ %ok
    Diopside & DD-MDD-074 (BKR1DD074) & 1.7 \\ %ok
    Fayalite & DD-MDD-098 (BKR1DD098) & 1.875 \\ %ok
    Ferrosaponite & JB-JLB-761-A (BKR1JB761A) & 1.58 \\ %ok
    %Fe-beidellite & BE-EAC-003 (LABE03) & 1.46 & Beidellite \\
    Forsterite & DD-MDD-085 (BKR1DD085) & 1.665 \\ %ok
    Gibbsite & HO-EAC-004-A (LAHO04A) & 1.6 \\ %ok
    Greenalite & GR-EAC-011-A (BKR1GR011A) & 1.685 \\ %ok
    Grossular & USGS, WS485 & 1.745 \\ %ok
    Hedenbergite & DL-CMP-082-A (BKR1DL082A) & 1.735 \\ %ok
    H$_2$O ice & Mastrapa et al.\cite{Mastrapa+2009} & given \\ %ok
    IOM & Kaplan et al.\cite{Kaplan+2018} & --$^{\rm a}$ \\ %ok
    Kaolinite & KA-EAC-001 (LAKA01) & 1.59 \\ %ok
    Laumontite & ZE-EAC-019 (LAZE19) & 1.525 \\
    Magnesite & JB-JLB-946 (BKR1JB946A) & 1.615 \\ %ok
    Magnetite & MG-EAC-002 (LAMG02) & 2.0$^{\rm b}$ \\ %ok
    %Mn(OH)$_2$ & -- & & No data available \\
    Minnesotaite & MN-EAC-001-A (BKR1MN001A) & 1.62 \\ %ok
    Muscovite & SR-JFM-071-A (BKR1SR071A) & 1.595  \\ %ok
    Pyrite & SC-EAC-119 (LASC119) & 5.0$^{\rm c}$ \\ %ok
    Quartz & JB-CMP-150 (995F150) & 1.545 \\ %ok
    Rhodochrosite & CB-EAC-068-A (BKR1CB068A) & 1.75 \\ %ok
    Saponite & SA-TXH-055 (LASA55) & 1.56 \\ %ok
    Siderite & CB-EAC-008-A (LACB08A) & 1.77 \\ %ok
    %Sylvite & -- & & Nodata available \\
    Talc & EA-EAC-015 (BKR1EA015) & 1.585 \\ %ok
    %Tephroite & OL-JMS-007 (C1OL07) & 1.875 & Limited to $<2.6\ \mu$m \\
    %Tremolite & CY-PLH-028 (C1CY28) & 1.64 & Limited to $<2.6\ \mu$m \\
    Troilite & EA-EAC-001-B (LAEA01B) & 4.0$^{\rm d}$  \\ %ok
    && \\
    %Saponite & SA-TXH-055 (LASA55) & Not used? \\
    %Daphnite & CH-EAC-016 (BKR1CH016) & Chamosite \\
    %Iron & SC-EAC-063 (LASC63) & Not used? \\
    %Pentlandite & AG-BXR-002 (LAAG02) & Not used? \\
    %Pyrrhotite & & \\
\end{tabular}
    \caption{References for endmember reflectances and refractive indices. Sample IDs are those in RELAB unless otherwise stated. The real part of refractive index $n$ is from  https://refractiveindex.info/ unless otherwise stated. a: Measured reflectance is used. b: \citeA{Glotch+Rossman2009}, c: \citeA{Sato+1984}, d: \citeA{Pollack+1994}.}
    \label{tabS:reflectance_data}
\end{table}

\newpage

\begin{table}[p]
    \centering
    \begin{tabular}{lr}
         Phase & Abundance [wt.\%]  \\ \hline
         SiO$_2$ & 36.314 \\
         Al$_2$O$_3$ & 3.598 \\
         FeO & 5.163 \\
         MnO & 0.204 \\
         MgO & 26.164 \\
         CaO & 2.893 \\
         Na$_2$O & 0.484 \\
         K$_2$O & 0.041 \\
         FeS & 6.563 \\
         Fe metal & 17.388 \\
         Pyrene & 1.116 \\
         HCl & 0.010 \\
         HCN & 0.063 \\ \hline
         Total & 100 \\
    \end{tabular}
    \caption{Initial bulk rock composition for geochemical modeling. A mean composition of CV chondrites was assumed \cite{Pearson+2006,Henderson+2009,Clay+2017}. S, Cl, N, and C were included as FeS, HCl, HCN, and pyrene plus HCN, respectively. The abundances of FeO and Fe metal were scaled to result in MgO/(MgO+FeO) = 0.9.}
    \label{tab:rock_composition}
\end{table}

\newpage

\begin{table}[p]
    \centering
    \begin{tabular}{ll}
    \hline
    Meteorite & ID or reference \\
    \hline
    Murchison (heated at 600$^\circ$C) & MT-JMS-190 (BKR1MT190) \\
    Cold Bokkeveld & \citeA{Ehlmann+2018} \\
                   & MT-JMS-186 (BKR1MT186) \\ 
    Tagish Lake &  MT-MEZ-011 (LCMT11)  \\
                &  MT-MEZ-012 (LCMT12) \\
    Ivuna (heated at 100$^\circ$C) & MP-TXH-018-F (LAMP18F) \\
    \hline
    \end{tabular}
    \caption{
    References for meteorites reflectance spectra. IDs are those in RELAB. Spectral data were selected and combined as reported in \citeA{Kurokawa+2020}.
    }
    \label{tabS:meteorite_data}
\end{table}

\newpage

\begin{table}[p]
    \centering
    \begin{tabular}{lllllll}
        \hline
        Name & Classification & Diameter [km] & Albedo & Semi-major axis [au] & Family & $3.1\ {\rm \mu m}$ abs. \\
        \hline
        1 Ceres & C & 973.9 & 0.087 & 2.767 & -- & A \\
        2 Pallas & B & 512.6 & 0.150 & 2.772 & F & -- \\
        10 Hygiea & C & 428.5 & 0.066 & 3.141 & F & A \\
        13 Egeria & Ch & 203.4 & 0.086 & 2.578 & -- & -- \\
        24 Themis & C & 176.8 & 0.084 & 3.137 & F & W \\
        49 Pales & Ch & 148.0 & 0.061 & 3.103 & -- & -- \\
        50 Virginia & Ch & 84.4 & 0.050 & 2.649 & -- & -- \\
        51 Nemausa & Cgh & 147.2 & 0.094 & 2.366 & -- & -- \\
        52 Europa & C & 350.4 & 0.043 & 3.093 & -- & A/W \\
        81 Terpsichore & C & 123.0 & 0.048 & 2.853 & F & A? \\
        94 Aurora & C & 179.2 & 0.053 & 3.156 & -- & A? \\
        106 Dione & Cgh & 153.4 & 0.084 & 3.180 & -- & -- \\
        121 Hermione & Ch & 194.1 & 0.058 & 3.448 & -- & -- \\
        128 Nemesis & C & 177.9 & 0.059 & 2.750 & F & A/W \\
        185 Eunike & C & 167.7 & 0.057 & 2.737 & -- & A \\
        308 Polyxo & T & 135.2 & 0.052 & 2.749 & -- & -- \\
        361 Bononia & D & 151.8 & 0.040 & 3.957 & -- & A \\
        423 Diotima & C & 226.9 & 0.049 & 3.067 & I & A \\
        451 Patientia & C & 234.9 & 0.071 & 3.061 & -- & A \\
        511 Davida & C & 291.0 & 0.070 & 3.164 & -- & A \\
        704 Interamnia & Cb & 316.2 & 0.075 & 3.059 & -- & A? \\
        \hline
    \end{tabular}
    \caption{A summary of asteroid properties. Classification: Bus-DeMeo taxonomy as summarized by \citeA{Hasegawa+2017}. Diameter, albedo, and semi-major axis: Data as summarized in \cite{Usui+2019}. Family: F (having a family), -- (not having a family), or I (interloper), as summarized by \citeA{Usui+2019,Vernazza+2021}. $3.1\ {\rm \mu m}$ absorption: A (ammoniated phyllosilicate), W (water ice), A/W (ammoniated phyllosilicates and/or water ice), and -- (non-detection). Possible detection (band depths are nonzero at their median values but 1-$\sigma$ error bars range to zero) is denoted by \textquotedblleft ?" at the end.}
    \label{tab:my_label}
\end{table}

\newpage

%Fig.1
\begin{landscape}
\begin{figure}
    \centering
    \includegraphics[width=1.0\linewidth]{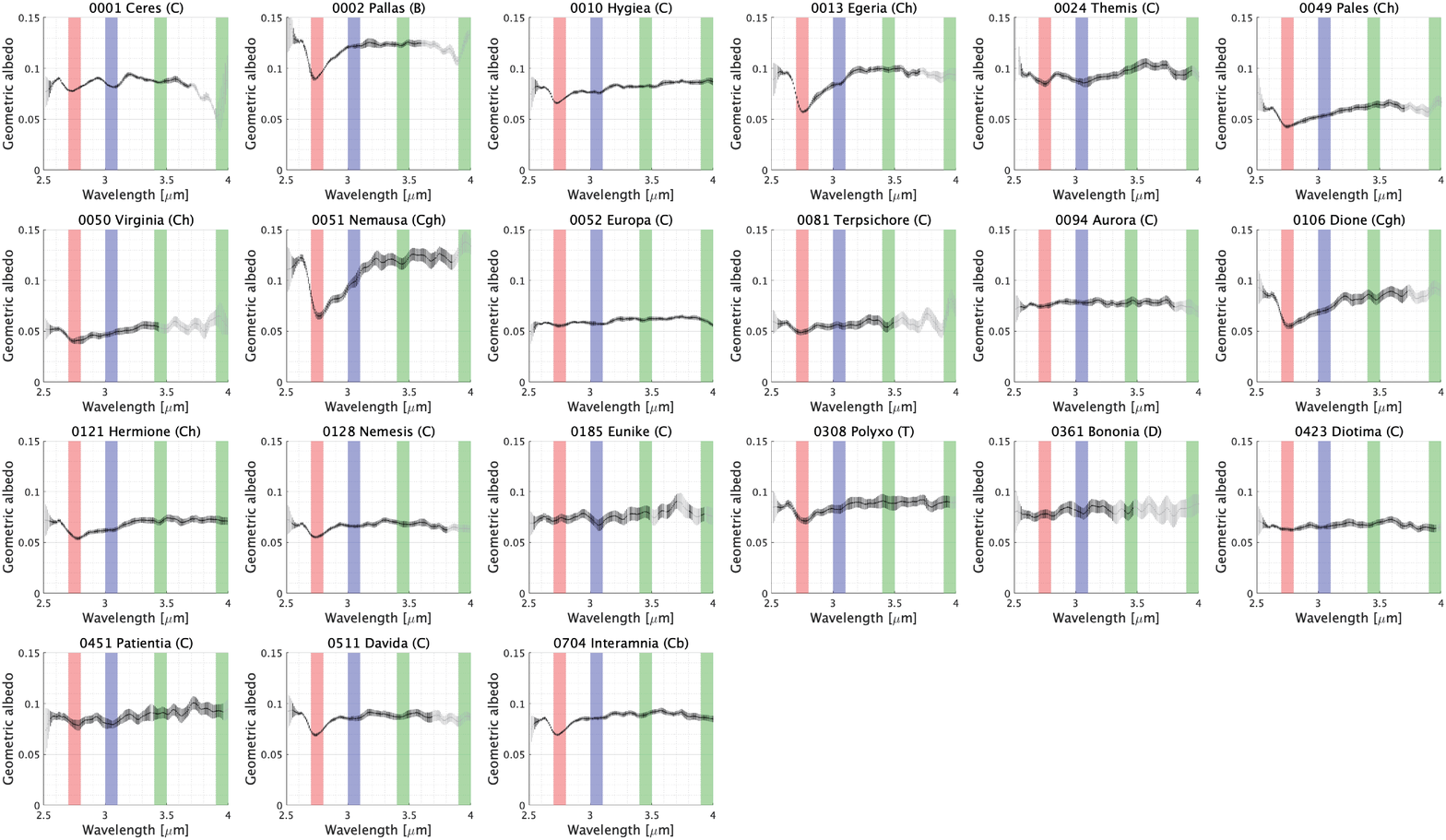}
    \caption{Reflectance spectra of the C-complex and D- and T-type asteroids observed by AKARI infrared space telescope. Data are from \citeA{Usui+2019}. The gray data points indicate the unreliable wavelength regions due to large uncertainties ($>10$\%) defined by \citeA{Usui+2019}. We highlighted the positions of absorption features: $2.7\ {\rm \mu m}$ (hydrous minerals, red), $3.1\ {\rm \mu m}$ (ammoniated saponite, blue), and $3.4$ and $4.0\ {\rm \mu m}$ (carbonates, green).}
    \label{figS:AKARI_spectra}
\end{figure}
\end{landscape}

\newpage

%Fig.2
\begin{landscape}
\begin{figure}
    \centering
    \includegraphics[width=1.0\linewidth]{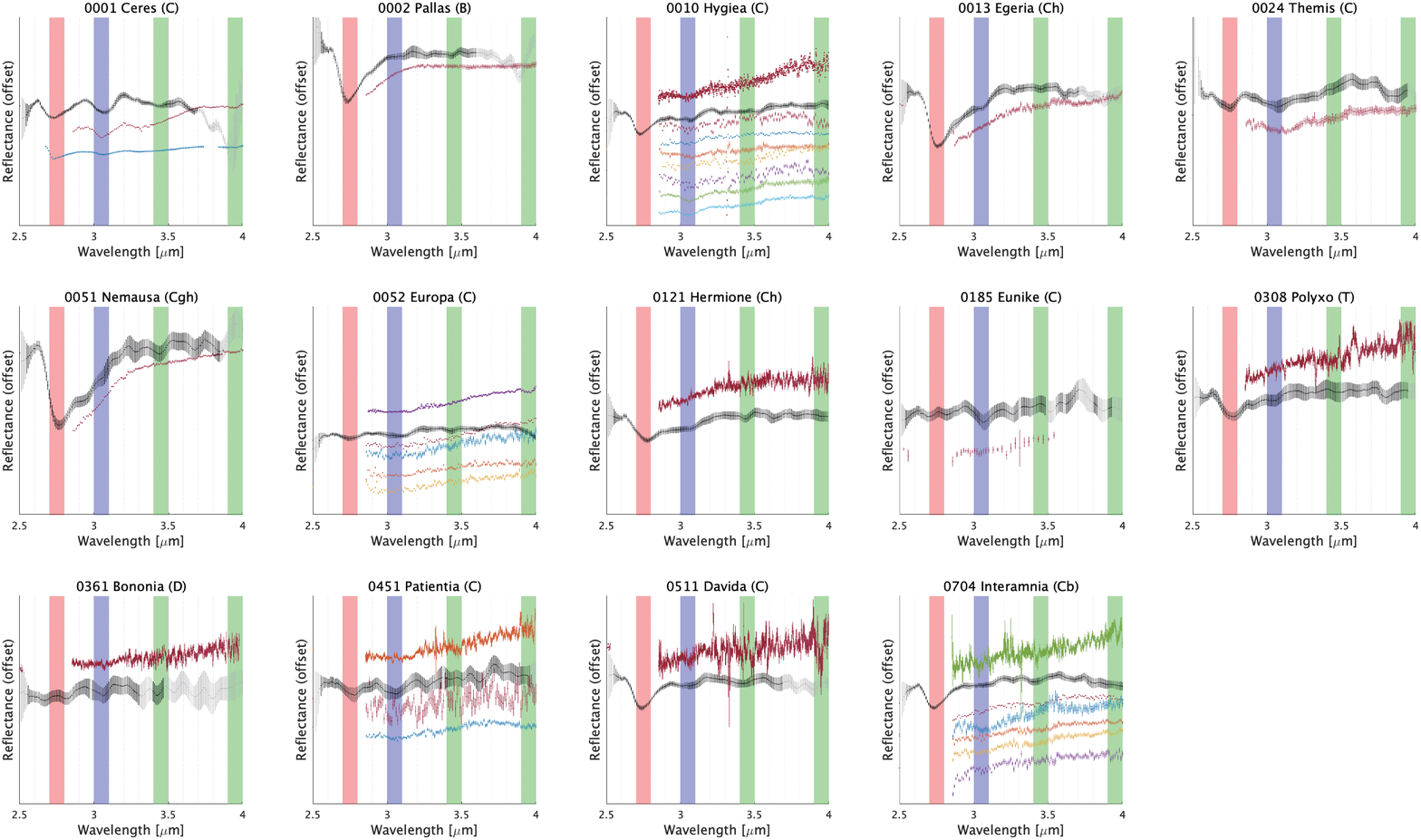}
    \caption{Comparison of reflectance spectra acquired using AKARI (black points), IRTF (colored points), and Dawn (blue points for Ceres).}
    \label{fig:AKARI_IRTF_DAWN_spectra}
\end{figure}
\end{landscape}

\newpage

%Fig.3
\begin{landscape}
\begin{figure}
    \centering
    \includegraphics[width=0.7\linewidth]{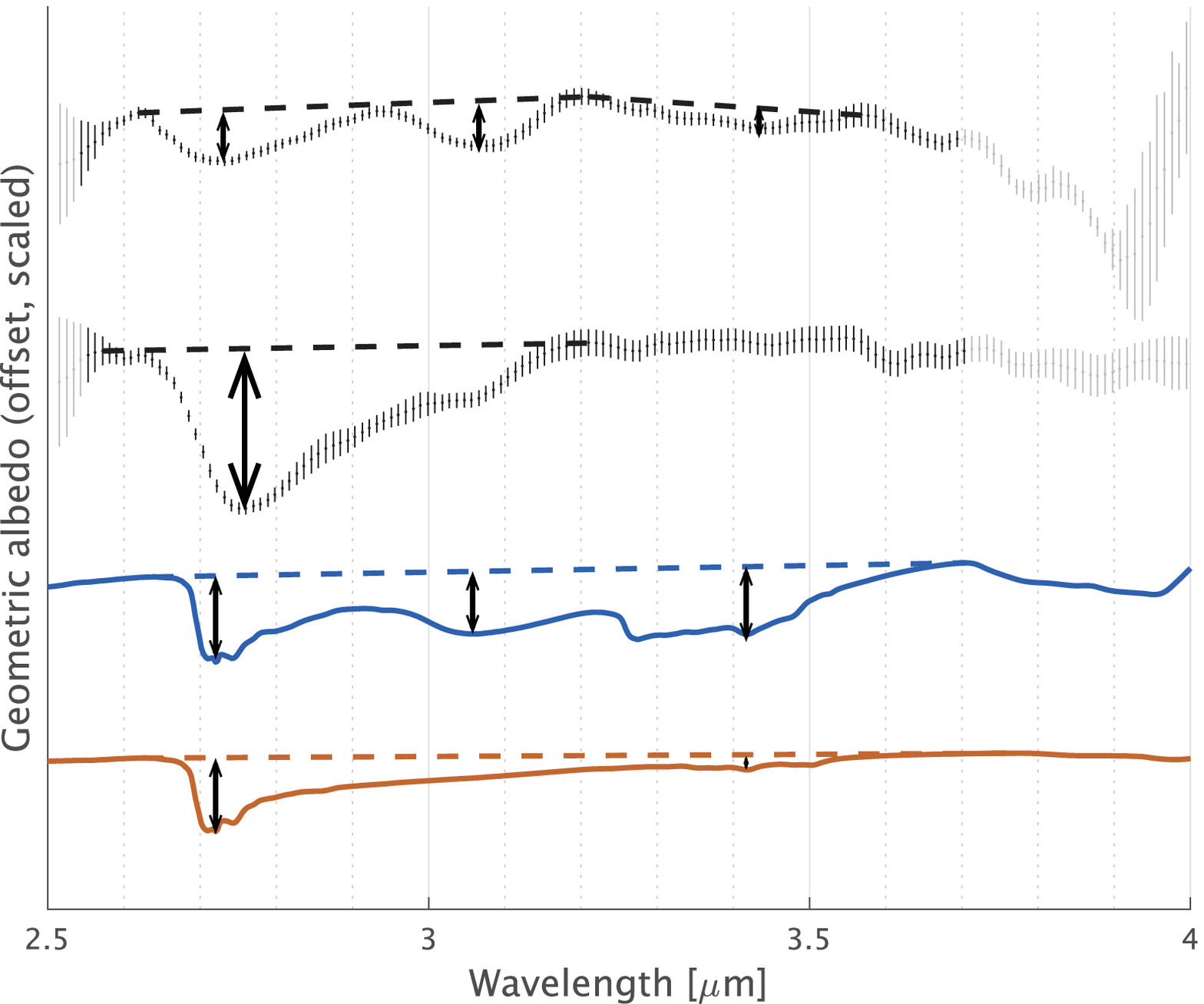}
    \caption{Definition of the absorption band depth. Black arrows indicate $R_c - R_\lambda$.}
    \label{figS:depth_definition}
\end{figure}
\end{landscape}

\newpage

%Fig.5
\begin{figure}
    \centering
    \includegraphics[width=1.0\linewidth]{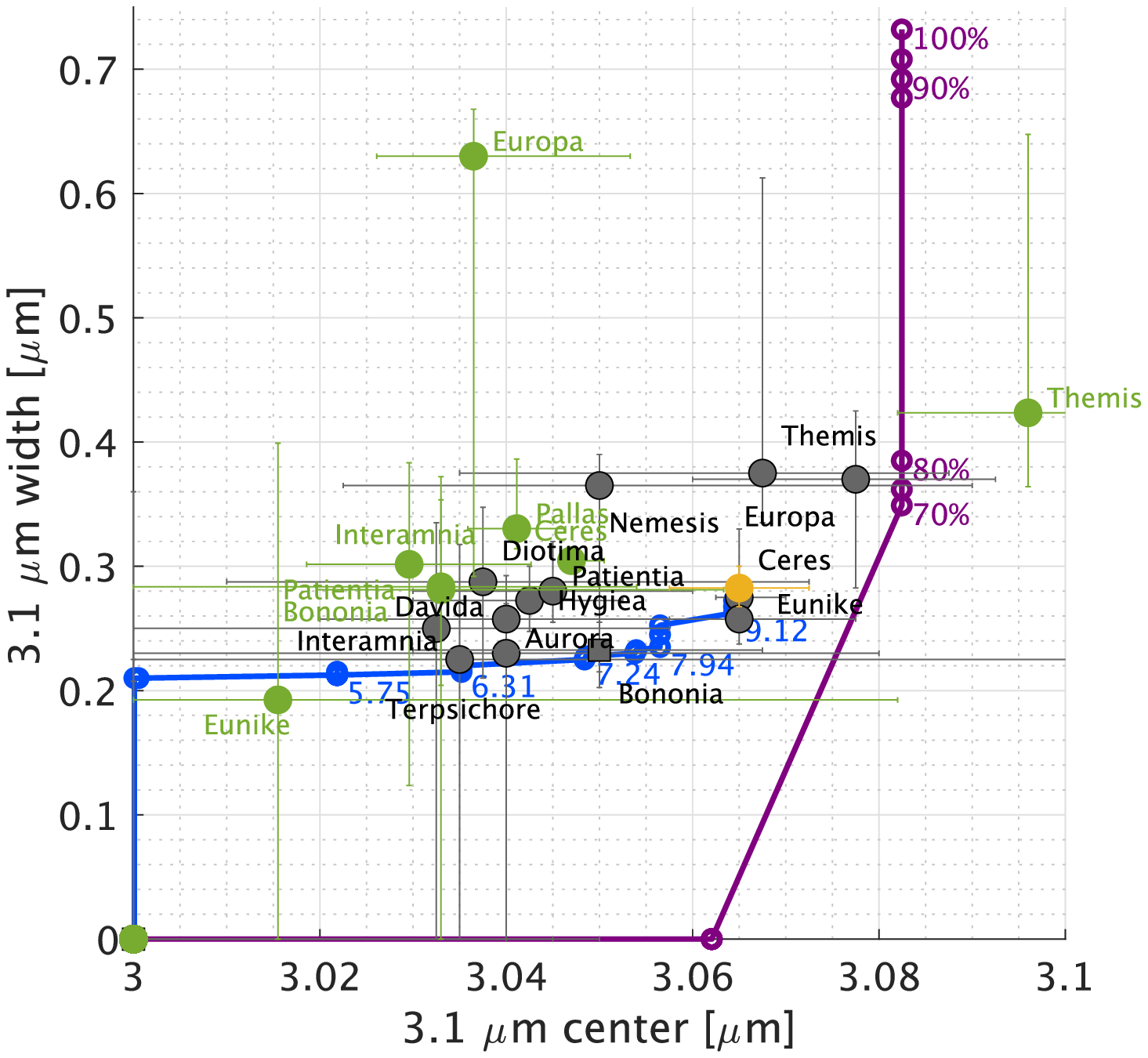}
    \caption{Center and width of the observed $3.1\ {\rm \mu m}$ absorptions for asteroids observed by AKARI (gray), IRTF (green), and Dawn for Ceres (yellow). Model lines are the same with those in Figure 2 in the main text.}
    \label{fig:Fig2separate}
\end{figure}

\newpage

%Fig.6
\begin{figure}
    \centering
    \includegraphics[width=1.0\linewidth]{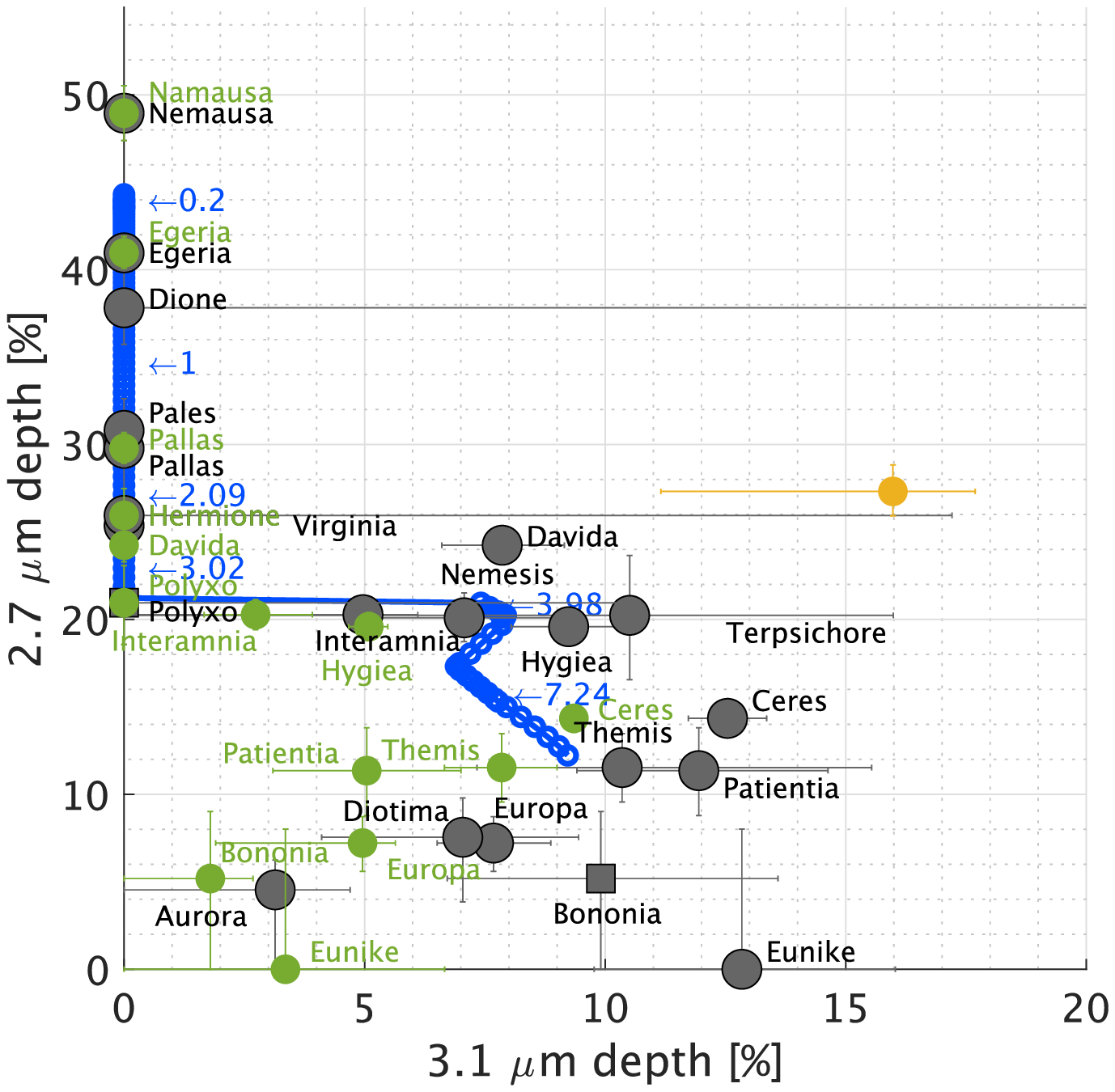}
    \caption{Absorption depths for asteroids observed by AKARI (gray), IRTF (green), and Dawn for Ceres (yellow). Model line is the same with that in Figure 3 in the main text.}
    \label{fig:features_34um}
\end{figure}

\newpage

%Fig.7
\begin{landscape}
\begin{figure}
    \centering
    \includegraphics[width=1.0\linewidth]{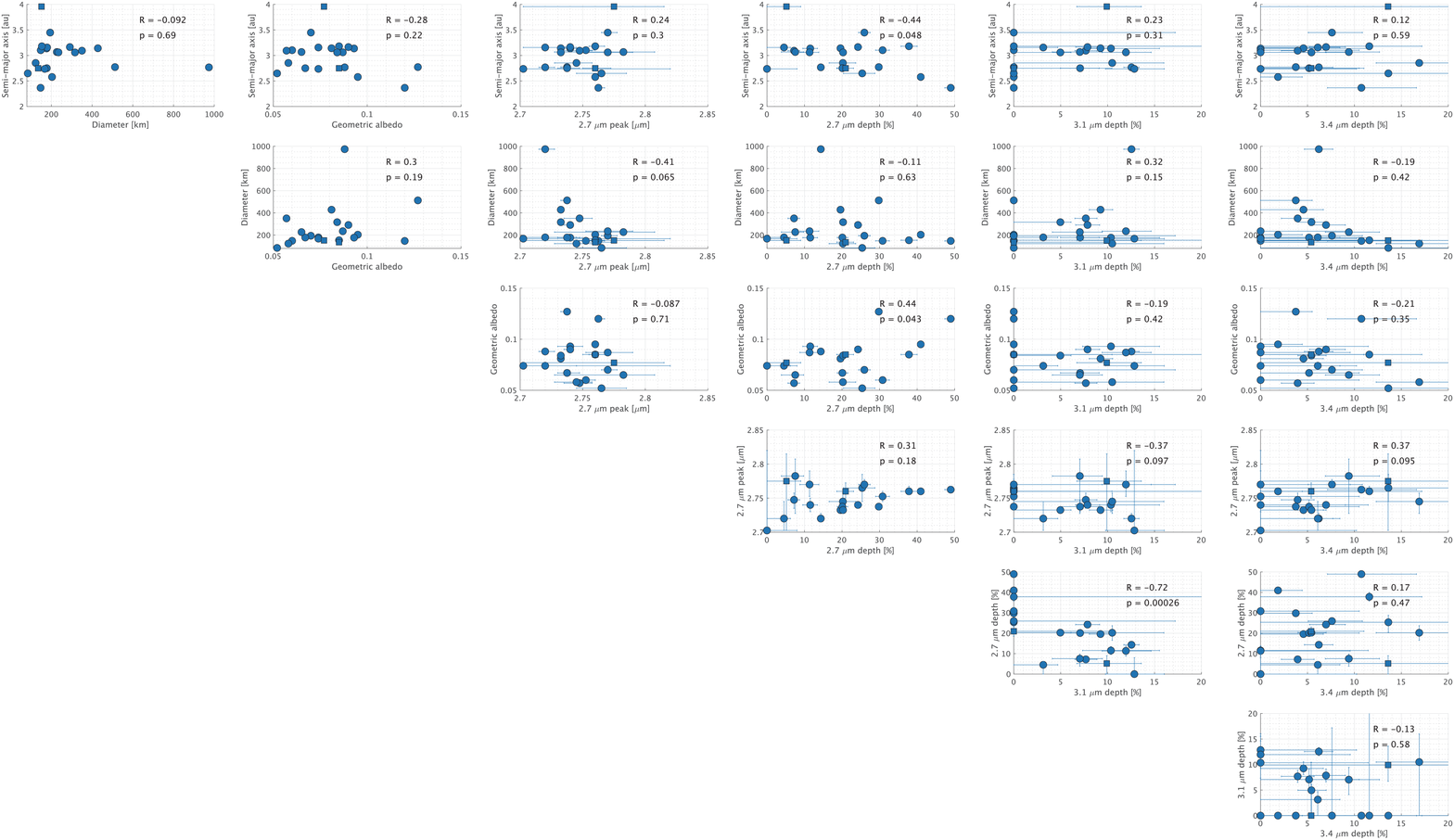}
    \caption{Correlation plot for the properties of C-complex (circle) and D- and T-type (square) asteroids observed by AKARI. The linear-correlation coefficient $R$ is given in each panel.}
    \label{figS:AKARI_correlation}
\end{figure}
\end{landscape}

\newpage

%Fig.8
\begin{landscape}
\begin{figure}
    \centering
    \includegraphics[width=1.0\linewidth]{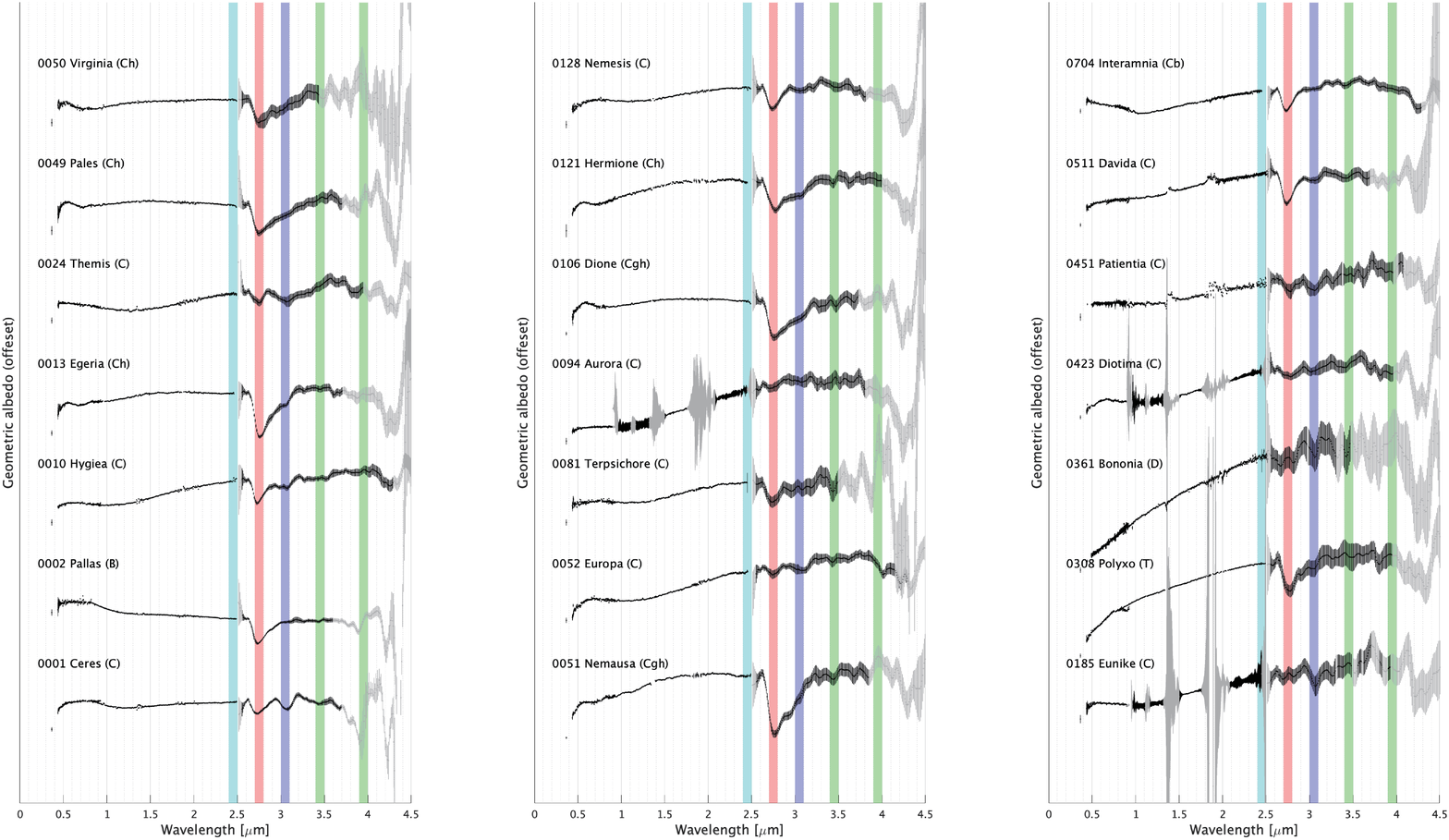}
    \caption{Combined spectra from visible to infrared wavelengths acquired using AKARI \cite{Usui+2019} and ground-based observations \cite{Bus+2002,Demeo+2009,Takir+Emery2012}. Data were compiled by \citeA{Hasegawa+2017}. The gray data points indicate the unreliable wavelength region due to large uncertainties ($>10$\%) defined by \citeA{Usui+2019}. We highlighted the positions of absorption features: $2.7\ {\rm \mu m}$ (hydrous minerals, red), $3.1\ {\rm \mu m}$ (ammoniated saponite, blue), $3.4$ and $4.0\ {\rm \mu m}$ (carbonates, green), and $2.45\ {\rm \mu m}$ (brucite, cyan).}
    \label{figS:AKARI+ground}
\end{figure}
\end{landscape}

\newpage

%Fig.9
\begin{landscape}
\begin{figure}
    \centering
    \includegraphics[width=1.0\linewidth]{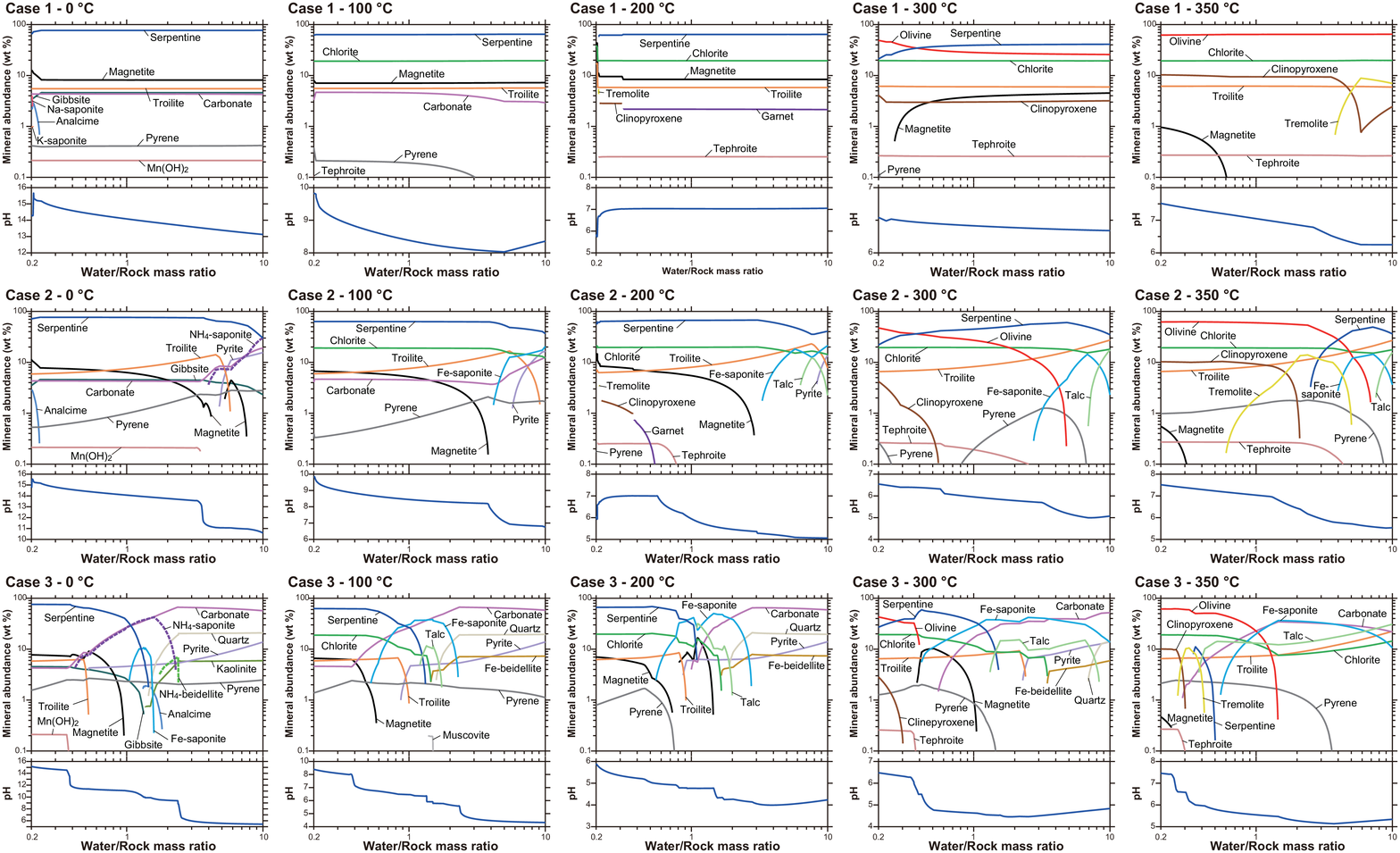}
    \caption{Mineral abundance and pH obtained from thermodynamic modeling as a function of W/R in Case 1--3 (rows from top to bottom) for $T = 0$, $100$, $200$, $300$, and $350^\circ$C (columns from left to right).}
    \label{figS:mineral_abundance}
\end{figure}
\end{landscape}

\newpage

%Fig.4
\begin{figure}
    \centering
    \includegraphics[width=1.0\linewidth]{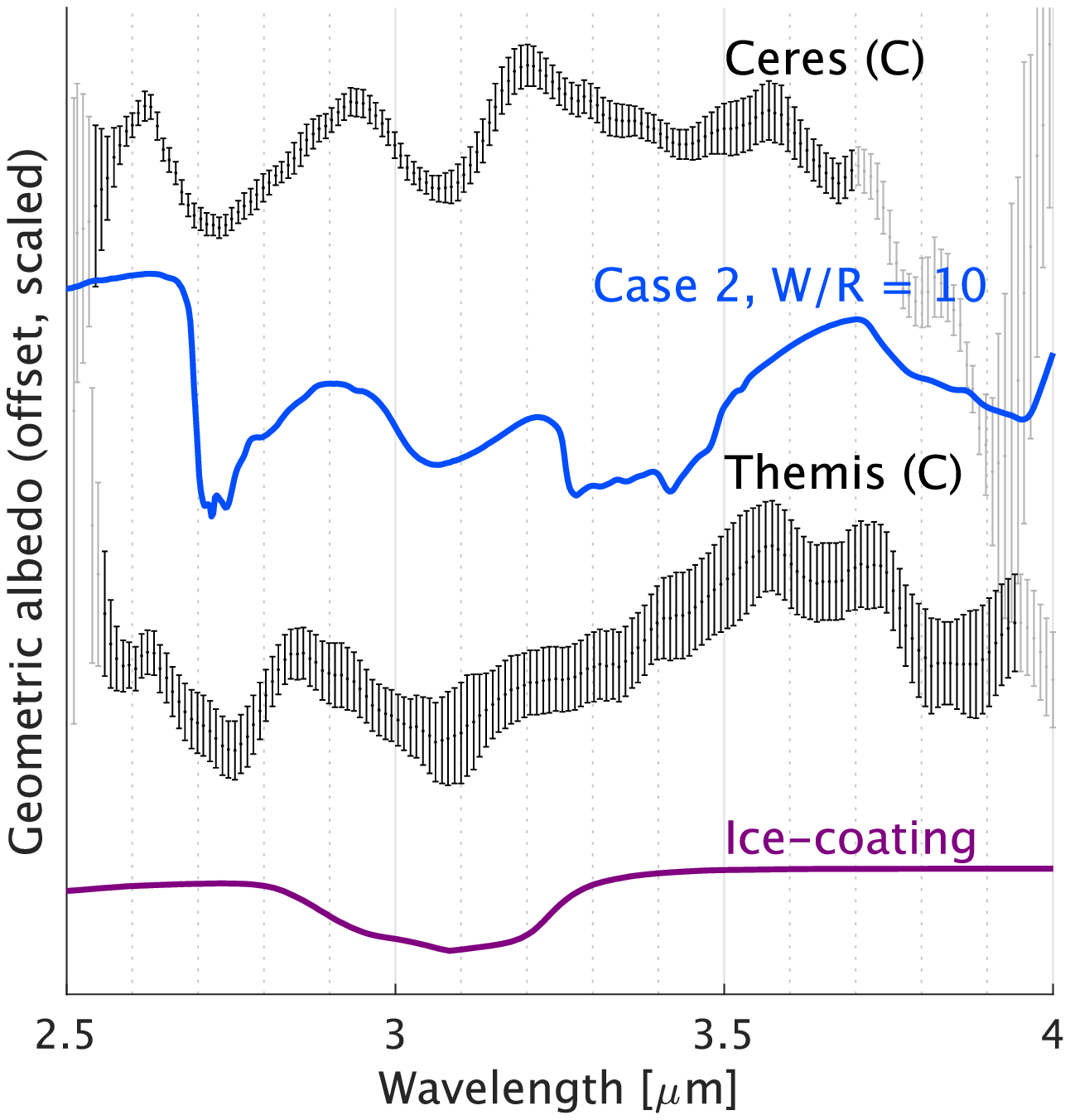}
    \caption{Comparison of asteroid spectra acquired using AKARI, Case 2 of our alteration model that we proposed creates asteroids with NH$_4$-bearing phyllosilicates, and the water-ice coating model of \citeA{Rivkin+Emery2010}. While 24 Themis is consistent with the water-ice coating model, 1 Ceres resembles the model with NH$_4$-bearing phyllosilicates.}
    \label{figS:31width}
\end{figure}

\newpage

%Fig.10
\begin{landscape}
\begin{figure}
    \centering
    \includegraphics[width=\linewidth]{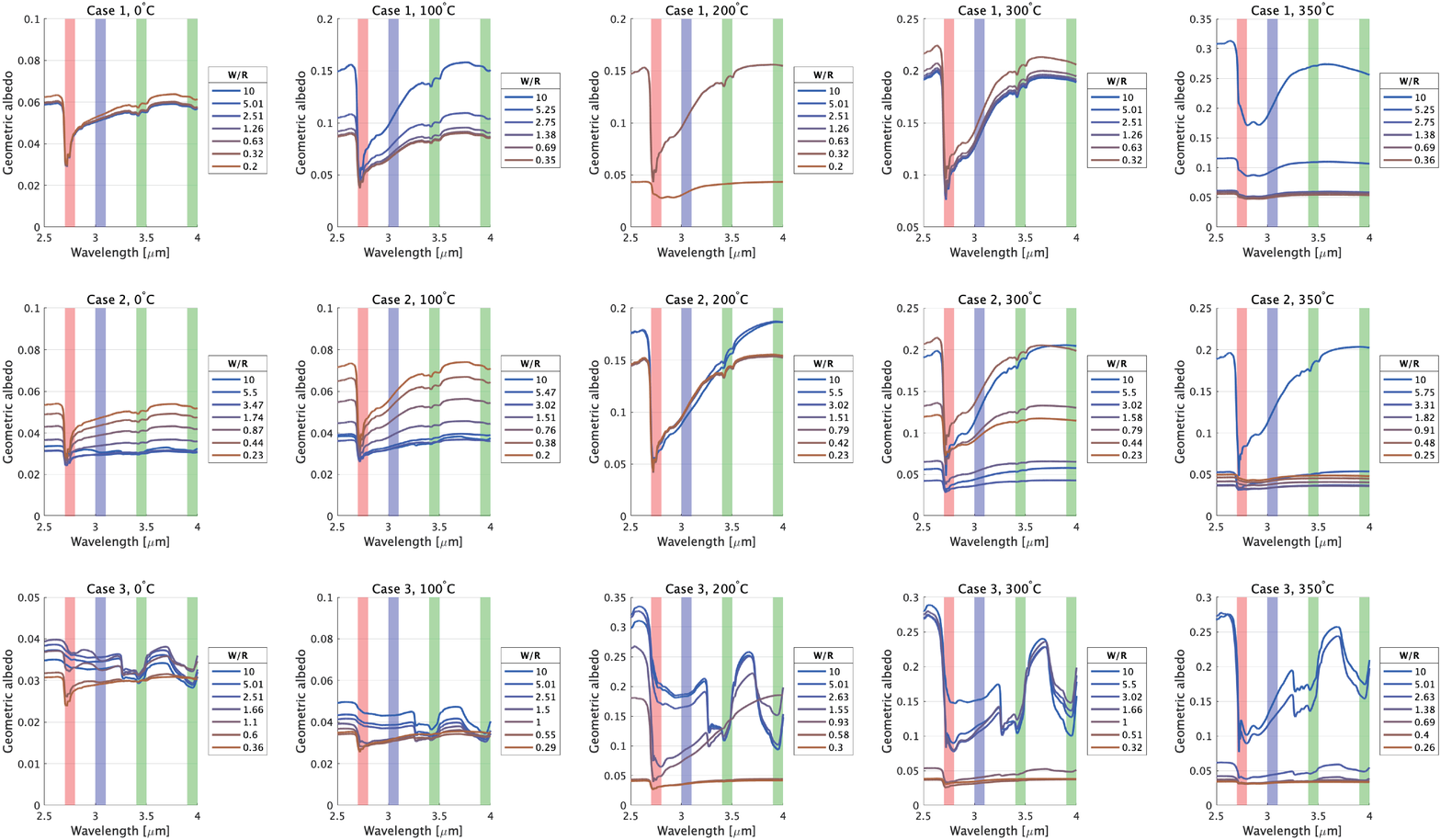}
    \caption{Model infrared reflectance spectra for various W/R in Case 1--3 (columns from left to right) and $T =$ 0, 100, 200, 300, and 350$^\circ$C (rows from top to bottom). We highlighted the positions of absorption features: $2.7\ {\rm \mu m}$ (hydrous minerals, red), $3.1\ {\rm \mu m}$ (ammoniated saponite, blue), and $3.4$ and $4.0\ {\rm \mu m}$ (carbonates, green).}
    \label{figS:all_spectra}
\end{figure}
\end{landscape}

\newpage

%Fig.11
\begin{figure}
    \centering
    \includegraphics[width=1.0\linewidth]{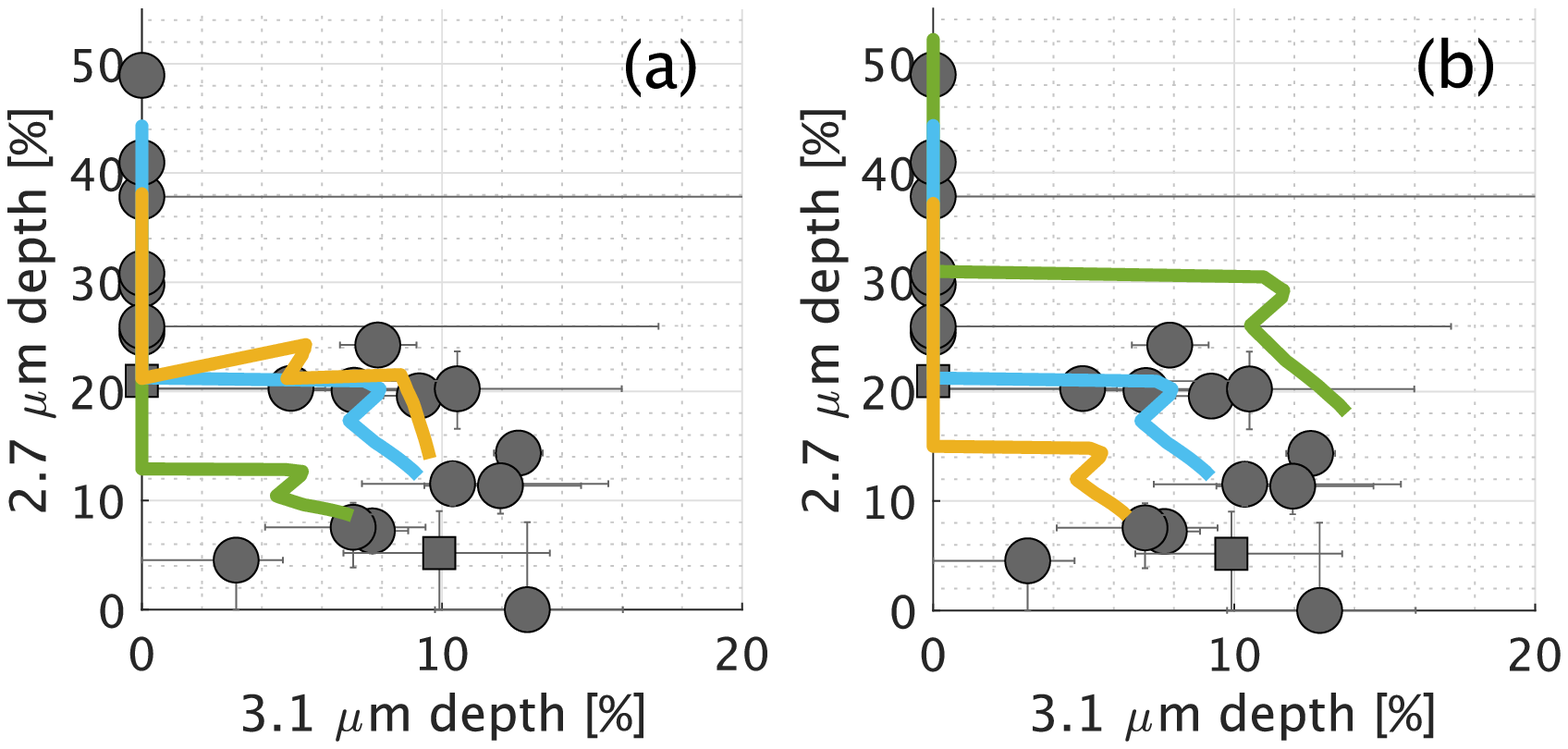}
    \caption{Dependence of modeled absorption band depths on grain sizes. (a) $d = 0.5\ {\rm \mu m}$ for IOM and $d = 200$ (green), $100$ (cyan), and $50\ {\rm \mu m}$ (yellow) for the other minerals. (b) $d = 100\ {\rm \mu m}$ for the other minerals and $d = 1$ (green), $0.5$ (cyan), and $0.3\ {\rm \mu m}$ (yellow) for IOM. Data points are the same with Figure 3 in the main text.}
    \label{figS:features_survey}
\end{figure}

\newpage

%Fig.12
\begin{landscape}
\begin{figure}
    \centering
    \includegraphics[width=0.7\linewidth]{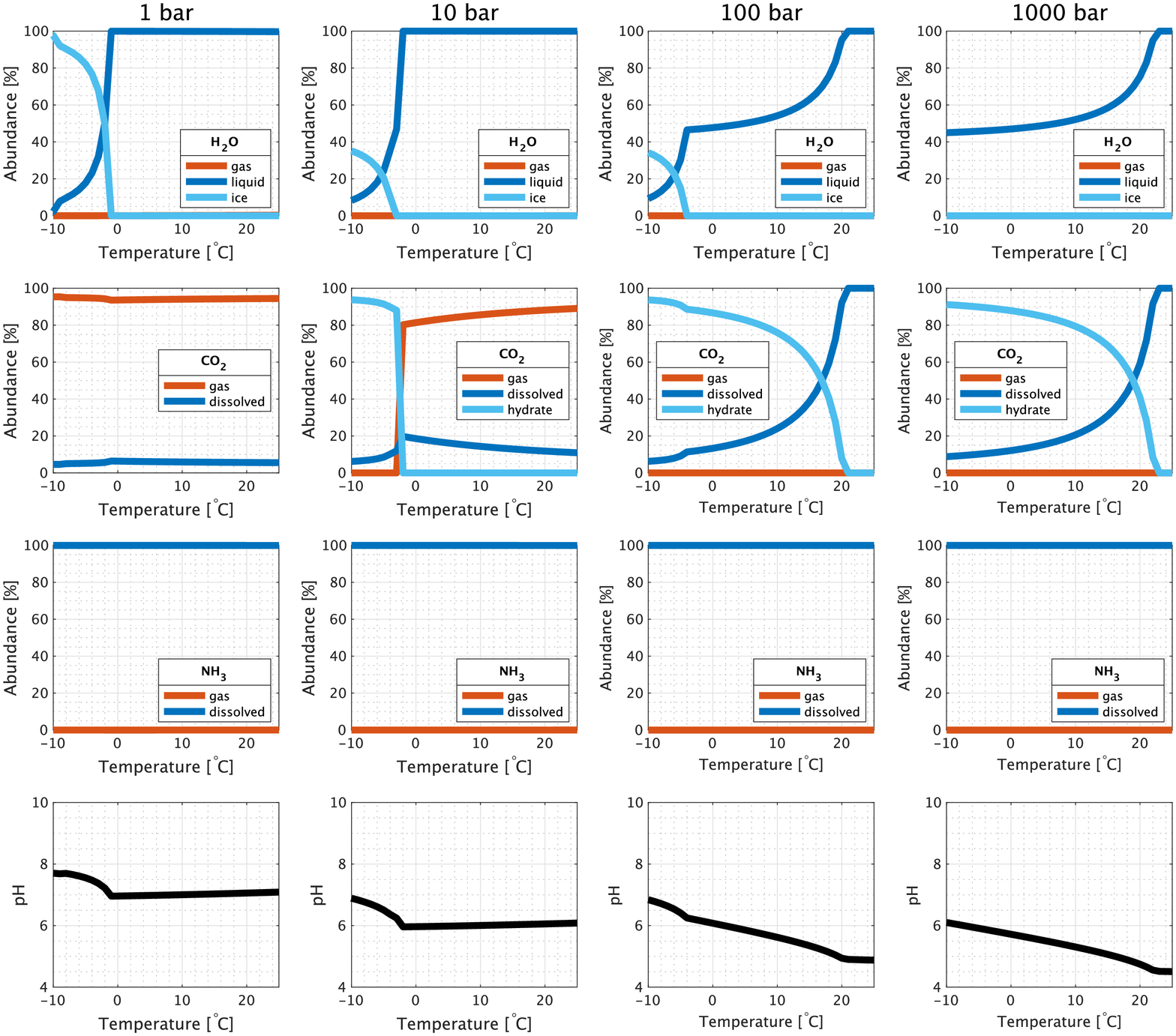}
    \caption{Abundances of gas, liquid, and solid phases for H$_2$O, CO$_2$, and NH$_3$  (rows from top to bottom) at 1, 10, 100, 1000 bar (columns from left to right) obtained from the thermodynamic calculations for the ice mixtures.}
    \label{figS:ice_mixture}
\end{figure}
\end{landscape}

\newpage

\acknowledgments
The data that support the plots {presented in} this paper are available at https://doi.org/10.6084/ m9.figshare.14338760.v1. AKARI data are available at http://www.ir.isas.jaxa.jp/AKARI/Archive/. 
EQ3/6 was developed by \citeA{Wolery+Jarek2003}. The code is available from Lawrence Livermore National Laboratory (LLNL).
GETFLOWS is a commercially-available hydrological simulator and is the property of Gesphere Environmental Technology Corp.
FREZCHEM is {an} open source code developed by \citeA{Marion+2012}. The code is available from G. Marion.
We thank the editor Francis Nimmo, Lucy F. Lim, and {three} anonymous reviewers for {their} careful review and constructive comments, which highly improved our manuscript.
We thank Andrew Rivkin, Driss Takir, Valerie Fox, and Hannah Kaplan for sharing reflectance data, Takehiro Hiroi and Moe Matsuoka for discussion about infrared reflectance spectra of meteorites, Mikhail Zolotov for sharing the modified version of the FREZCHEM code, and Keisuke Fukushi for providing thermodynamic parameters for smectites.
This study was supported by JSPS KAKENHI Grant number 15K05277, 17H01175, 17H06454, 17H06455, 17H06456, 17H06457, 17H06458, 17H06459, 17K05636, 18K13602, 19H00725, 19H01960, 19H05072, 20KK0080, {21H04514,} 21K13976, and JSPS Core-to-Core Program "International Network of Planetary Sciences."
Part of the data utilized in this publication were obtained and made available by the MITHNEOS MIT-Hawaii Near-Earth Object Spectroscopic Survey. The IRTF is operated by the University of Hawaii under Cooperative Agreement no. NCC 5-538 with the National Aeronautics and Space Administration, Office of Space Science, Planetary Astronomy Program. The MIT component of this work is supported by NASA grant 09-NEOO009-0001, and by the National Science Foundation under Grants Nos. 0506716 and 0907766.

%% ------------------------------------------------------------------------ %%
%% References and Citations

%%%%%%%%%%%%%%%%%%%%%%%%%%%%%%%%%%%%%%%%%%%%%%%
%
% \bibliography{<name of your .bib file>} don't specify the file extension
%
% don't specify bibliographystyle
%%%%%%%%%%%%%%%%%%%%%%%%%%%%%%%%%%%%%%%%%%%%%%%

%\newpage

\bibliography{sample.bib}

%Reference citation instructions and examples:
%
% Please use ONLY \cite and \citeA for reference citations.
% \cite for parenthetical references
% ...as shown in recent studies (Simpson et al., 2019)
% \citeA for in-text citations
% ...Simpson et al. (2019) have shown...
%
%
%...as shown by \citeA{jskilby}.
%...as shown by \citeA{lewin76}, \citeA{carson86}, \citeA{bartoldy02}, and \citeA{rinaldi03}.
%...has been shown \cite{jskilbye}.
%...has been shown \cite{lewin76,carson86,bartoldy02,rinaldi03}.
%... \cite <i.e.>[]{lewin76,carson86,bartoldy02,rinaldi03}.
%...has been shown by \cite <e.g.,>[and others]{lewin76}.
%
% apacite uses < > for prenotes and [ ] for postnotes
% DO NOT use other cite commands (e.g., \citet, \citep, \citeyear, \nocite, \citealp, etc.).
%

\end{document}